\documentclass[aps, pra, reprint, showpacs, amsmath, amssymb, superscriptaddress, 10pt]{revtex4-1}
\usepackage{graphicx}
\usepackage{bm}
\usepackage{epsfig}
\usepackage{amsmath}

\def\be{\begin{equation}}
\def\ee{\end{equation}}
\def\bea{\begin{eqnarray}}
\def\eea{\end{eqnarray}}

\begin{document}
\title{Heralding efficiency and correlated-mode\\ coupling of near-IR fiber coupled photon pairs}

\author{P.\ Ben Dixon}
\affiliation{Lincoln Laboratory, Massachusetts Institute of Technology, Lexington, MA 02420, USA}
\affiliation{Research Laboratory of Electronics, Massachusetts Institute of Technology, Cambridge, MA 02139, USA}

\author{Danna Rosenberg}
\affiliation{Lincoln Laboratory, Massachusetts Institute of Technology, Lexington, MA 02420, USA}

\author{Veronika Stelmakh}
\affiliation{Research Laboratory of Electronics, Massachusetts Institute of Technology, Cambridge, MA 02139, USA}

\author{Matthew E.\ Grein}
\affiliation{Lincoln Laboratory, Massachusetts Institute of Technology, Lexington, MA 02420, USA}

\author{Ryan S.\ Bennink}
\affiliation{Oak Ridge National Laboratory, Oak Ridge, Tennessee 37831, USA}

\author{Eric A.\ Dauler}
\affiliation{Lincoln Laboratory, Massachusetts Institute of Technology, Lexington, MA 02420, USA}

\author{Andrew J.\ Kerman}
\affiliation{Lincoln Laboratory, Massachusetts Institute of Technology, Lexington, MA 02420, USA}

\author{Richard J.\ Molnar}
\affiliation{Lincoln Laboratory, Massachusetts Institute of Technology, Lexington, MA 02420, USA}

\author{Franco N.\ C.\ Wong}
\affiliation{Research Laboratory of Electronics, Massachusetts Institute of Technology, Cambridge, MA 02139, USA}

\date{\today}

\begin{abstract}
We report on a systematic experimental study of heralding efficiency and generation rate of telecom-band infrared photon pairs generated by spontaneous parametric down-conversion and coupled to single mode optical fibers.  We define the correlated-mode coupling efficiency---an inherent source efficiency---and explain its relation to heralding efficiency.  For our experiment, we developed a reconfigurable computer controlled pump-beam and collection-mode optical apparatus which we used to measure the generation rate and correlated-mode coupling  efficiency.  The use of low-noise, high-efficiency superconducting nanowire single-photon detectors in this setup allowed us to explore focus configurations with low overall photon flux.  The measured data agree well with theory and we demonstrated a correlated-mode coupling efficiency of \( 97 \pm 2\% \), which is the highest efficiency yet achieved for this type of system.  These results confirm theoretical treatments and demonstrate that very high overall heralding efficiencies can, in principle, be achieved in quantum optical systems.  It is expected that these results and techniques will be widely incorporated into future systems that require, or benefit from, a high heralding efficiency.
\end{abstract}


\maketitle
The development of quantum enabled technologies promises substantial benefit for applications such as increased computing power and enhanced cryptographic security \cite{knill_scheme_2001, kok_linear_2007, aaronson_computational_2011, ralph_quantum_2013, bennett_quantum_1984, ekert_quantum_1991, hughes_refining_2011, weedbrook_gaussian_2012}. The continued development of this field, however, faces significant obstacles.  Photonic systems can generate high quality quantum states; but there is difficulty in efficiently detecting these states.  One important photonic quantum state is the entangled photon pair state.  A source of these photon pair states can be used as a heralded single photon source or, more importantly, it can be used in a Bell test measurement that ensures quantum behavior of the photon pair \cite{*[{}][{ in Chap.~3.}] Sakurai_Modern_1993, Bell_EPR_1964, Brunner_Bell_2014}.  This type of use has been the focus of significant experimental effort recently because, with very efficient detection, it promises to enable Bell test measurements that are free from both the detection and locality loopholes \cite{wittmann_loophole-free_2012, christensen_detection-loophole-free_2013, giustina_bell_2013}.  These loopholes may be exploited---either naturally, or by a malicious attack---to make a classical system incorrectly appear to be acting in a quantum mechanical manner.  Closing the loopholes requires extraordinary effort, thus placing many of the benefits to quantum technologies that are afforded by Bell measurements, as well as clarity on fundamental and philosophical considerations regarding the quantum nature of reality, beyond current reach \cite{eberhard_background_1993, weihs_violation_1998}. 

In an ideal photon pair experiment one would detect both photons from the pair.  However, system inefficiencies necessarily result in the loss of one or both of the photons from the pair.  Detecting both photons remains challenging even with state of the art superconducting single-photon detectors \cite{rosenberg_high-speed_2013, marsili_detecting_2013, miki_high_2013} because the relevant characteristic is not simply the detector efficiency, but rather the more demanding total system efficiency.  This total system efficiency is comprised of three component efficiencies: the detector efficiency \( \eta_{d} \), the transmission efficiency of the optical system \(\eta_{s} \), and what we call the correlated-mode coupling efficiency \( \eta_{c} \).  This correlated-mode coupling efficiency can be thought of as the inherent source efficiency; it is the probability that one photon from the pair (or a noise photon from the source) couples to a detector, given that its pair photon (or, again, a noise photon from the source) has been coupled to a detector.  Because \( \eta_{c} \) is a conditional probability and deals with system-dependent multi-mode photon pair emission structure, it is more complex than the other component efficiencies.  There has been comparatively little, but nevertheless promising, research specifically investigating this effect \cite{ljunggren_optimal_2005, mosley_heralded_2008, bennink_optimal_2010, cunha_pereira_demonstrating_2013, Guerreiro_HighEfficiency_2013}.  A better understanding and characterization of a quantum system's \(\eta_{c} \) is then crucial to the development of future quantum technologies and experimental investigations of quantum foundations.

Here we report on a systematic experimental study of \( \eta_{c} \) in one of the most commonly used systems to generate entangled photons: a bulk nonlinear optical crystal generating photon pairs through the process of spontaneous parametric down-conversion (SPDC), with coupling into single mode optical fibers.  We developed a reconfigurable computer controlled pump-beam and collection-mode optical apparatus which we used to measure the optimized \( \eta_{c} \) over a range of pump-beam parameters and collection-mode parameters.  The use of low-noise, high-efficiency superconducting nanowire single-photon detectors (SNSPDs) in this setup allowed us to explore focus configurations with low overall photon flux to reduce multi-pair events, while still getting reliable measurements of \( \eta_{c} \) \cite{rosenberg_high-speed_2013}.  The measured photon detection rates and \( \eta_{c} \) agree well with the theoretical treatment of Bennink \cite{bennink_optimal_2010}.  Moreover, we demonstrated a correlated-mode coupling efficiency of \( 97\% \), which is the highest efficiency yet achieved for this type of system.  These results confirm that very high \( \eta_{c} \) can be achieved by quantum optical systems---indicating that very high overall heralding efficiencies can, in principle, be achieved.  It is expected that these results and techniques to optimize \( \eta_{c} \) will be widely incorporated into photonic systems that benefit from a high heralding efficiency, such as loophole-free Bell test systems or heralded single photon sources \cite{shields_semiconductor_2007, pomarico_mhz_2012}.

\section{Correlated-Mode Coupling Efficiency}
The correlated-mode coupling can be understood generally for multi-mode, two particle quantum states, which very closely approximates the SPDC output \cite{*[{}][{ in Chap.~22.}] mandel_optical_1995, hong_experimental_1986}.  This multi-mode two particle quantum state can be written as:
\begin{equation}
|\Psi\rangle = \sum\limits_{n=0}^{\infty}\sum\limits_{m=0}^{\infty}\psi(m,n) \hat{a}^{\dagger}_{m} \hat{b}^{\dagger}_{n} |0\rangle,
\label{State1}
\end{equation}
where \(\hat{a}^{\dagger}\) and \(\hat{b}^{\dagger}\) are creation operators for the two output particles, \(a\) and \(b\) (commonly called signal and idler photons for SPDC).  The index \(m\) enumerates the modes available to particle \(a\) and \(n\) enumerates the modes available to particle \(b\).  The function \(\psi(m,n)\) is the mode probability amplitude for the two particle state.  

We can associate the single collected modes of particles \(a\) and \(b\)  with modes \(m=0\) and \(n=0\), respectively.  Relevant, mutually exclusive mode emission probabilities are the following: 
\begin{align}
        P_{p} &= \left|\langle 0|\hat{a}_{0}\hat{b}_{0}|\Psi\rangle\right|^{2} = \left| \psi(0,0) \right|^{2}, \label{ProbPKet}\\ 
\tilde{P}_{a} &= \left| \langle 0 | \sum\limits_{n=1}^{\infty} \hat{a}_{0}\hat{b}_{n} | \Psi \rangle\right|^{2} = \left|  \sum\limits_{n=1}^{\infty} \psi(0,n) \right|^{2}, \label{ProbAKet}\\ 
\tilde{P}_{b} &= \left|\langle 0 |  \sum\limits_{m=1}^{\infty} \hat{a}_{m}\hat{b}_{0}|\Psi\rangle\right|^{2} = \left|  \sum\limits_{m=1}^{\infty} \psi(m,0) \right|^{2}, \label{ProbBKet}
\end{align}
where \(P_{p}\) is the probability of both particles \(a\) and \(b\) emitted into the collected modes---meaning they would both be detected if a perfect collection and detection system were used.  \(\tilde{P}_{a}\) and \(\tilde{P}_{b}\) are the probabilities of only particle \(a\) or only particle \(b\), respectively, emitted into the collected modes---meaning the partner particle is necessarily lost.  

The total probability for particle \(a\) to be emitted into the collected mode (independent of its partner) is then \(P_{a} = \tilde{P}_{a} + P_{p}\).  Similarly, the total probability for particle \(b\) to be emitted into the collected mode is \(P_{b} = \tilde{P}_{b} + P_{p}\)

A family of related conditional mode coupling efficiencies can then be defined depending on how the conditioning is done.  There is the probability that particle \(a\) is emitted into the collected mode, conditioned on particle \(b\) being emitted into the collected mode---given by the formula \(P_{p}/P_{b}\).  Similarly, there is the probability that particle \(b\) is emitted into the collected mode, conditioned on particle \(a\) being emitted into the collected mode---given by the formula \(P_{p}/P_{a}\).  Finally, there is the symmetrically conditioned multiplicative average of these probabilities which we call the correlated-mode coupling efficiency \(\eta_{c}\):
\begin{equation}
\eta_{c} = \frac{P_{p}}{\sqrt{P_{a} P_{b}}}.
\label{EtaC1}
\end{equation}
Because \(\eta_{c}\) is symmetric with respect to particles \(a\) and \(b\), optimizing this metric yields a system with high coupling efficiencies for both particles.  This is more useful because both particles have the desired coupling characteristics, although it is more difficult to realize experimentally. 

It is important to distinguish these mode coupling efficiencies, which are properties of the quantum state alone, from true heralding efficiencies, which depend upon details of the experiment.  

A family of heralding efficiencies can be defined depending, again, on how conditioning is done. The heralding efficiency for particle \(a\), with conditioning on particle \(b\) is defined as the probability that particle \(a\) is emitted into the collected mode, conditioned on particle \(b\) being collected and detected (described by \(P_{p}/P_{b}\)), and that particle \(a\), once emitted into the collected mode, is transmitted or otherwise made available for some purpose (described by efficiency \(\tilde{\eta}_s\)).  We see then that the heralding efficiency for particle \(a\) is \( P_{p}/P_{b} \times \tilde{\eta}_s \).  The heralding efficiency for particle \(b\), conditioned on detecting particle \(a\), can be similarly defined, with the roles of \(a\) and \(b\) interchanged, and the symmetric heralding efficiency is the multiplicative average of these two efficiencies.

In these heralding efficiency formulas, the efficiency that the heralded particle is made available for some purpose \(\tilde{\eta}_s\) is defined operationally, depending on the specific application.  For example, if the application is to simply transmit the heralded particle to a specific location, \(\tilde{\eta}_s\) would be the efficiency of transmission. Alternatively, if the application is analyzing detection statistics (such as for a Bell test measurement), the particle is only available for this purpose if it is ultimately detected, meaning that  \(\tilde{\eta}_s\) for this case is the product of the transmission efficiency to the detector and the detection efficiency of the detector itself.

\section{Quantum State}
To determine the functional form of \(\eta_{c}\) for our SPDC system, we follow the theoretical treatment of Bennink \cite{bennink_optimal_2010}.  We model the pump beam as having a Gaussian spatial profile focused to the center of the nonlinear crystal with waist \(w_{p}\) and wavenumber \(k_{p} = n_{p} \omega_{p}/c\), where \(n_{p}\) is the index of refraction the pump beam experiences inside the crystal, \(\omega_p\) is the optical frequency of the pump beam, and \(c\) is the speed of light in vacuum.  The pump mode focal parameter is then defined as \(\xi_{p} = L / k_{p} w_{p}^{2}\), where \(L\) is the length of the nonlinear crystal. 

We similarly model the collected modes of photon \(a\) and photon \(b\) as having Gaussian profiles focused at the center of the nonlinear crystal, both with waist \(w_{c}\), and wavenumbers \(k_{a} = n_{a} \omega_{a}/ c \) and \(k_{b} = n_{b} \omega_{b}/ c\), respectively, where \(n_{a}\) and \(n_{b}\) are the respective indices of refraction experienced inside the crystal, \(\omega_a\) and \(\omega_b\) are the respective optical frequencies.  The focal parameters for the collected modes of photon \(a\) and photon \(b\) are defined, respectively, as \(\xi_{a} = L / k_{a} w_{a}^{2} \) and \(\xi_{b} = L / k_{b} w_{b}^{2} \).  We model the pump mode and collected \(a\) and \(b\) modes as mutually collinear and all propagating along a crystal principal axis.  The optical fields experience wavenumber mismatch \(\Delta k = \left(k_{p} - k_{a} - k_{b}\right)\) inside the crystal and we model the crystal as having poling period \(\Lambda\) that brings the nonlinear interaction into the quasi-phase-matching regime, whose strength is given by the effective nonlinearity of the crystal \(d_{\mathrm{eff}}\).  

For this system, the emission rate of photon \(a\) into the collected mode per milliwatt of pump light \(R_{a}\), the emission rate of photon \(b\) into the collected mode per milliwatt of pump light \(R_{b}\), and the emission rate of both photons from the correlated pair into the collected mode per milliwatt of pump light \(R_{c}\) can be approximated as follows \cite{bennink_optimal_2010, DixonBennink_emails_2014}:
\begin{align}
R_{a} &= \frac{128 \pi^{2} \lambda_p}{10^3 \epsilon_{0}  n_{p}^2 |n_{a}' - n_{b}'|}   \left(\frac{d_{\mathrm{eff}} }{\lambda_{a} \lambda_{b}}\right)^{2} \frac{\arctan\left( \frac{ B_{a} }{ A_{a} } \xi_{a} \right) }{ A_{a} B_{a}}, \label{ProbAFormula}\\
R_{b} &= \frac{128 \pi^{2} \lambda_p}{10^3 \epsilon_{0}  n_{p}^2 |n_{a}' - n_{b}'|}  \left(\frac{d_{\mathrm{eff}} }{\lambda_{a} \lambda_{b}}\right)^{2} \frac{\arctan\left( \frac{ B_{b} }{ A_{b} } \xi_{b} \right) }{ A_{b} B_{b}}, \label{ProbBFormula}\\
R_{c} &= \frac{128 \pi^{2} \lambda_p}{10^3 \epsilon_{0}  n_{p}^2 |n_{a}' - n_{b}'|}  \left(\frac{d_{\mathrm{eff}} }{\lambda_{a} \lambda_{b}}\right)^{2} \frac{\arctan\left( \frac{ B_{+} }{ A_{+} } \frac{\xi_{a} \xi_{b}}{\xi_{p}} \right) }{ A_{+} B_{+}}. \label{CRateTH}
\end{align}
In these equations \( n_{a}' = c\, \partial k_{a} / \partial \omega_{a} \), \( n_{b}' = c\, \partial k_{b} / \partial \omega_{b} \), and
\begin{align}
A_{a} &= 2 \sqrt{\left( 1+\frac{ k_{a} }{ k_{p} } \frac{\xi_{a}}{\xi_{p}} \right) \frac{k_{b}}{k_{p}}  },  \label{AaFormula}\\
A_{+} &=  1 + \frac{ k_{a} }{ k_{p} } \frac{\xi_{a}}{\xi_{p}} + \frac{ k_{b} }{ k_{p} } \frac{\xi_{b}}{\xi_{p}},  \label{AplusFormula}\\
B_{a} &= 2\left( 1-\frac{\Delta k}{k_{p}} \right) \sqrt{\left( 1+\frac{ k_{a} + \Delta k }{ k_{p} - \Delta k } \frac{\xi_{a}}{\xi_{p}} \right) \frac{k_{b} + \Delta k}{k_{p}-\Delta k} }, \label{BaFormula}\\
B_{+} &=  \left( 1-\frac{\Delta k}{k_{p}} \right) \left( 1 + \frac{ k_{a} + \Delta k }{ k_{p} - \Delta k } \frac{\xi_{p}}{\xi_{a}} + \frac{ k_{b} + \Delta k }{ k_{p} - \Delta k } \frac{\xi_{p}}{\xi_{b}} \right). \label{BplusFormula}
\end{align}

Although these equations were formulated specifically for the conditions in the experiment reported here, the equations are generally applicable and accurate to about \(20\%\) or better under common conditions, namely when: emission is in the visible or near IR; the crystal is not very short ($L \gtrsim 1 \operatorname{mm}$); the phase mismatch is small ($\Delta k \ll k_a, k_b, k_p$); the group velocity dispersion is negligible; and the focusing is not extremely tight ($\xi_a,\xi_b,\xi_p \lesssim 10$).  The interested reader may refer to \cite{bennink_optimal_2010} for a fuller discussion of the approximations underlying these formulas and the impact on their accuracy.

Formulas for \(A_{b}\) and \(B_{b}\) can be found by interchanging indices \(a\) and \(b\) in the formulas for \(A_{a}\) and \(B_{a}\), respectively.  The total photon emission rate is:
\begin{equation}
R_{t} = R_{a} + R_{b}.
\label{TotalRateTH}
\end{equation}

These rates represent the relevant probabilities multiplied by an overall generation rate. Thus we can substitute Eqs.~(\ref{ProbAFormula})--(\ref{CRateTH}) into Eq.~(\ref{EtaC1}) to express the correlated-mode coupling efficiency in terms of experimentally determined parameters:
\begin{equation}
\eta_{c} =  \frac{\sqrt{A_{a} B_{a} A_{b} B_{b}}  \arctan\left( \frac{B_{+}}{A_{+}} \frac{\xi_{a} \xi_{b}}{\xi_{p}} \right)} {A_{+} B_{+}  \sqrt{\arctan\left( \frac{B_{a}}{A_{a}} \xi_{a} \right)  \arctan\left( \frac{B_{b}}{A_{b}} \xi_{b} \right)} }.
\label{EtaTH}
\end{equation}

\begin{figure}
\includegraphics[scale=0.80]{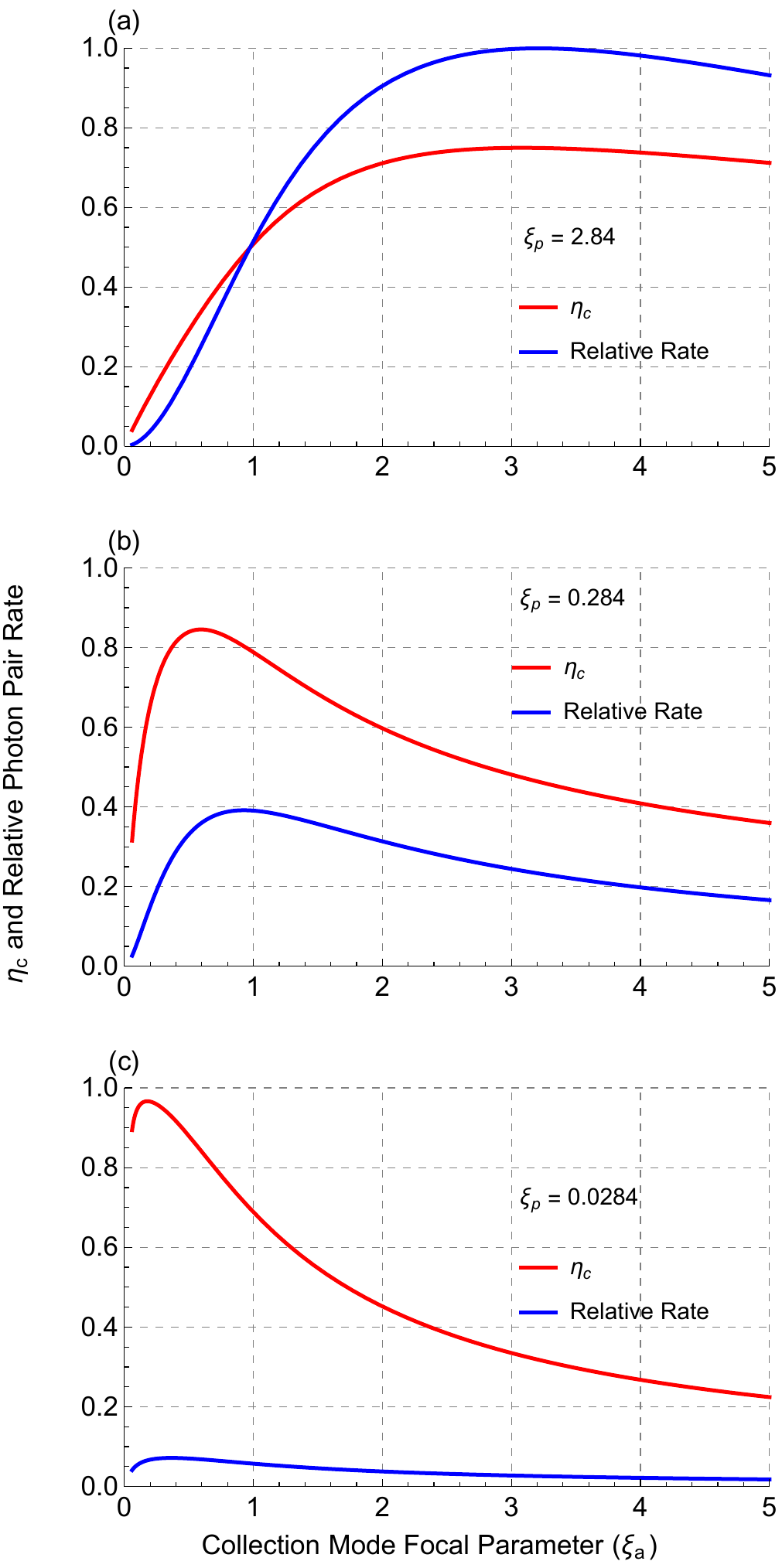}
\caption{\label{TheoryPlot1}(color online) Correlated-mode coupling behavior and relative photon pair emission rate as a function of collection mode focal parameter \(\xi_{a}\) for a range of pump beam focal parameters \(\xi_{p}\) are shown. Plot (a) shows the behavior for \(\xi_{p} = 2.84\), all photon pair rates in this figure are normalized to the maximum pair rate achieved for this focusing configuration. Plot (b) shows the behavior for \(\xi_{p} = 0.284\), and plot (c) shows the behavior for \(\xi_{p} = 0.0284\).}
\end{figure} 
The pair emission rate and correlated-mode coupling behavior for several pump beam focal parameters are shown in Fig.~\ref{TheoryPlot1}.  In these plots we use wavelength and crystal parameters corresponding to those from the experiment described in this manuscript. Figure \ref{TheoryPlot1}(a) shows the behavior for \(\xi_{p} = 2.84\).  Because this pump focal parameter is generally considered to be a good choice for efficient second harmonic generation, we use the peak pair emission rate for this focus as our rate normalization baseline  \cite{*[{}][{ in Chap.~2.}] Boyd_nonlinear_2008, boyd_parametric_1968}.  We see that although this focal parameter is generally considered to be a good choice, its correlated-mode coupling efficiency has an upper limit of \(\sim\!\!75\%\).  Figures \ref{TheoryPlot1}(b) and \ref{TheoryPlot1}(c) show that by moving to looser pump focusing conditions, we can dramatically increase the correlated-mode coupling efficiency; however, this benefit comes at the expense of decreased correlated pair emission rate.  To wit, for \(\xi_{p} = 0.0284\) we can achieve a correlated-mode coupling efficiency in excess of \(96 \%\), albeit with a normalized pair rate of only \(7\%\).  It is this loose pump focusing regime---with the promise of very high correlated-mode coupling efficiency---that we focus on.

\section{Experimental Configuration}
\begin{figure}
\includegraphics[scale=0.75]{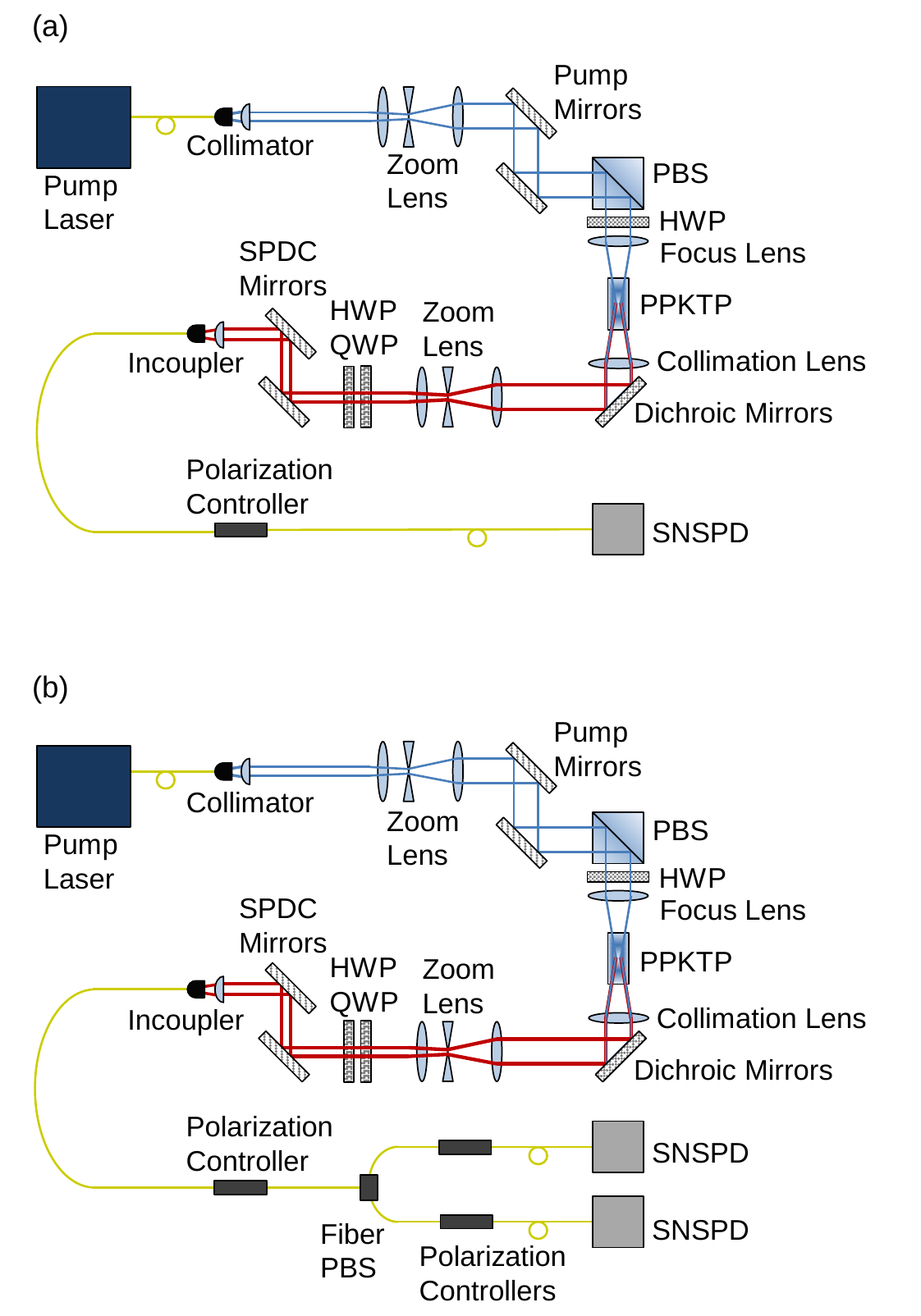}
\caption{\label{Expt}(color online) The experimental configuration for the single detector apparatus (a) and for the two detector apparatuses (b) are shown.  For both configurations the pump beam is launched into free space, sent through a variable zoom lens and computer controlled steering mirrors, and then focused to the center of the PPKTP crystal.  The crystal generates photon pairs which pass through a collimation lens, reflect off four dichroic mirrors, pass through a computer controlled zoom lens, and are then coupled into a single mode optical fiber. For configuration (a), the photons pass through a fiber polarization controller and into the SNSPD apparatus. For configuration (b), the orthogonally polarized photon pairs pass through a fiber polarization controller, are separated at a fiber polarizing beam splitter, sent through fiber polarization controllers and into the SNSPD apparatuses.  For each configuration, the detector outputs are sent to timing and counting hardware and recorded on a computer.}
\end{figure}

Our two experimental setups are shown in Fig.~\ref{Expt};  for both cases we used a 40\,mW, fiber coupled, CW laser with 780\,nm wavelength as the pump beam.  A fiber beam collimator launched this beam into free space, resulting in a nearly Gaussian profile beam which we sent through a reconfigurable pump beam zoom lens, giving collimated beams of various sizes.  We then finely aligned the pump beam with a pair of computer controlled steering mirrors and tuned the beam polarization with a half wave plate.  We focused the beam with a 300\,mm focal length lens to the center of a 1\,cm long periodically-poled potassium titanyl phosphate (PPKTP) crystal.  The combination of the focusing lens and the reconfigurable zoom lens allowed for a broad range of pump mode focal parameters \(\xi_{p}\).  

The PPKTP crystal had a \(46.1\,\mu \)m poling period, designed to generate orthogonally polarized correlated photon pairs at 1560\,nm from energy degenerate, collinear, type II phase-matched spontaneous parametric down-conversion (SPDC).  The relevant indices of refraction for this crystal were determined from the set of Sellmeier equations in references \cite{Fradkin_Tunable_1999} and \cite{Konig_Extended_2004}.  We sent the SPDC output through a second \(f = 300\)\,mm lens, which we called the collimation lens, located 1 focal length from the crystal that limited the spread of the multimode SPDC output.  We reflected the SPDC output off four dichroic mirrors which filtered residual pump beam light and directed the SPDC output through a reconfigurable, computer controlled zoom lens.  A fiber incoupler then coupled the SPDC photons into a single mode optical fiber.  

The computer controlled SPDC zoom lens, in conjunction with the optical fiber acted as a spatial filter, allowing a single spatial mode---the collection mode---to be sorted out of the multimode SPDC output.  This spatial mode was characterized by the collection mode focal parameters \(\xi_{a}\) and \(\xi_{b}\), however because both the \(a\) and \(b\) photons were collected with the same optical setup, the mode focal parameters differed only by which birefringent crystal index was as a multiplicative factor: \(n_{a}\) or \(n_{b}\).  Our experimental configuration allowed for collection mode focal parameters ranging from \(\xi_{a} = 0.1\) to \(\xi_{a} = 0.7\).  

The fiber coupled SPDC was then sent to one of two detection systems.  The initial detection setup, shown in Fig.~\ref{Expt}(a), made use of a single detector apparatus. In this configuration we sent the fiber coupled SPDC through a fiber polarization controller and then directly to a detector apparatus consisting of four interleaved SNSPDs with independent outputs \cite{rosenberg_high-speed_2013}.  The polarization controller was used to maximize the coincident photon detection by the polarization dependent SNSPDs.  However, because the SPDC photon pairs were orthogonally polarized, neither the \(a\) photons nor the \(b\) photons were detected with optimum efficiency in this configuration.  

The four detector outputs were then sent to a PicoQuant HydraHarp 400 event timer which interfaced with computer software to record the total detection rate \(\mathfrak{R}_{t}\), the coincident photon detection rate \(\mathfrak{R}_{c}\) and their ratio.  \(\mathfrak{R}_{t}\) was found by summing the detection events from all channels, and \(\mathfrak{R}_{c}\) was found by summing twofold detection events between any two channels within a 1\,ns coincidence window.  These events were averaged over 1\,s integration times to determine rates.

The second detection system, shown in Fig.~\ref{Expt}(b), made use of two separate detector apparatuses which allowed for more direct data analysis.  In this configuration we aligned the polarization of the fiber coupled SPDC output using a fiber polarization controller and then deterministically split the orthogonally polarized \(a\) and \(b\) photons with a fiber polarizing beam splitter.  We then routed the \(a\) and \(b\) photons into separate system arms.  In both of these arms, we sent the component SPDC outputs through a fiber polarization controller to optimize the polarization and then into a detector apparatus with four interleaved SNSPDs.  

The outputs of these two detector apparatuses (4 outputs from each apparatus) were then sent to an 8 channel PicoQuant HydraHarp 400 interfacing with computer software which recorded the total photon \(a\) detection rate \(\mathfrak{R}_{a}\), the total photon \(b\) detection rate \(\mathfrak{R}_{b}\), the coincident photon detection rate between the arms \(\mathfrak{R}_{c}\), and real time ratios of coincident to total detection rates.  The total photon detection rate for each arm was found by summing the 4 outputs of the arm's SNSPD apparatus.  The coincident photon detection rate was found by summing twofold detection events between any channel from one arm with any channel from the other arm, again within a 1\,ns coincidence window. These events were again averaged over 1\,s integration times to determine rates.

The detector apparatuses consisted of four separate 80 nm wide niobium nitride wires arranged in a 14\,\(\mu\)m wide interleaved pattern.  These SNSPDs were cooled to 3\,K in a housing with a single optical fiber input.  This type of nanowire detector is known to have efficiencies that can exceed 75\% at 1560\,nm, with 5\,ns reset time, and dark counts that can be brought below 1000\,counts per second (cps) \cite{rosenberg_high-speed_2013}.  These beneficial detector characteristics gave the novel capability to accurately and quickly measure the performance of our optical system in focus configurations ranging from those with high total count rates but low coincident count rates, to configurations with low total count rates with high \(\eta_c\) values.  

\section{System Considerations}
In order to find the inherent source emission rates and \(\eta_{c}\) values we must divide out the effects of detector efficiency, detector dark counts, and optical transmission efficiency.  However, we must take care to account for these effects correctly, especially for the configuration with only one detector apparatus because blocking effects are then present.  Single photon dark count rates \(D\) and coincident photon dark count rates \(D_{c}\) can be determined by accurately characterizing the total system.

Event probabilities for the case of a pair of coincident photons collected by a single SNSPD apparatus are given in Table \ref{2PhotonTable}.  These probabilities are given in terms of  the single photon transmission efficiency of the optical system that routes the photon to the SNSPD \(\eta_s\) and the single photon detection efficiency of the SNSPD \(\eta_d\).  

An important consideration is that the derivation of these formulas makes use of the simplifying assumption that each individual nanowire has equal efficiency \(\eta_d/4\). The true component nanowire efficiencies were observed to be slightly different (all component efficiencies were within \(1\%\) of the mean component efficiency), however accounting for the differing efficiencies results in a correction to the terms in Table \ref{2PhotonTable} that is much less than \(1\%\), indicating that the simplifying assumption is appropriate.    
\begin{table}
\setlength{\tabcolsep}{8pt}
\caption{\label{2PhotonTable} Detection possibilities for coincident photons collected by a single SNSPD apparatus are shown.  Detection possibilities, the symbols used to denote them, and their associated probabilities, in terms of component system efficiencies, are listed for an SNSPD apparatus collecting SPDC photons.}
\begin{ruledtabular}
\begin{tabular}{l|l|l}
Possibilities & Symbol & Probability  \\  \hline
Two detection events &\(P(2|2)\) &  \( 3 (\eta_s \eta_d)^{2}/4 \) \\
One detection events &\(P(1|2)\) &  \( 2\eta_s \eta_d - 7(\eta_s \eta_d)^{2}/4 \) \\
No detection events&\(P(0|2)\) &  \( 1-2\eta_s \eta_d+(\eta_s \eta_d)^{2} \)
\end{tabular}
\end{ruledtabular}
\end{table}

Event probabilities for the case of only a single photon collected by an SNSPD apparatus are given in Table \ref{1PhotonTable}.  These probabilities are again given in terms of \(\eta_s\) and \(\eta_d\), and are not affected by differing component nanowire efficiencies.
\begin{table}
\setlength{\tabcolsep}{10pt}
\caption{\label{1PhotonTable} Detection possibilities for a single photon collected by a single SNSPD apparatus are shown.  Detection possibilities, the symbols used to denote them, and their associated probabilities, in terms of component system efficiencies, are listed for an SNSPD apparatus collecting at most one of the SPDC photons.}
\begin{ruledtabular}
\begin{tabular}{l|l|l}
Possibilities & Symbol & Probability  \\  \hline
One detection events &\(P(1|1)\) &  \( \eta_s \eta_d  \) \\
No detection events&\(P(0|1)\) &  \(  1-\eta_s \eta_d  \)
\end{tabular}
\end{ruledtabular}
\end{table}

\subsection{Single SNSPD Configuration}
In the experimental configuration consisting of a single SNSPD apparatus, shown in Fig.~\ref{Expt}(a), individual rates of the \(a\) and \(b\) photons cannot be distinguished, only the total photon detection rate \(\mathfrak{R}_{t}\) and the coincident photon pair detection rate \(\mathfrak{R}_{c}\) can be measured.  Additionally, the single and coincident photon dark count rates \(D\) and \(D_{c}\) can be measured (\(D_{c}\) includes coincidences between two dark counts as well as coincidences between a single dark count and a single detected \(a\) or \(b\) photon).  We can subtract the detector dark count rates to determine the corrected total photon detection rate \(\tilde{\mathfrak{R}}_t = \mathfrak{R}_t - D\) and the corrected coincident photon rate \(\tilde{\mathfrak{R}}_c = \mathfrak{R}_c - D_{c}\).  Expressions for these detector dark count corrected detection rates are as follows:
\begin{align}
\tilde{\mathfrak{R}}_{t} = &R_{c}\left( P(1|2) + 2 P(2|2) \right) \nonumber \\
								 &+ \left(R_{a} + R_{b} - 2 R_{c}\right) P(1|1), \label{TotalDetectedRate1}\\
\tilde{\mathfrak{R}}_{c} = &R_{c} P(2|2). \label{PairDetectedRate1}
\end{align}
We can invert these equations and substitute expressions from Tables \ref{2PhotonTable} and \ref{1PhotonTable} to express the true emission rates and \(\eta_{c}\) in terms of \(\eta_d\), \(\eta_s\) and the measured detection rates:
\begin{align}
R_{c} &= 4  \tilde{\mathfrak{R}}_{c} / 3 (\eta_s \eta_d)^{2}, \label{CRateEXP1}\\
R_{t} &= \left( 3 \tilde{\mathfrak{R}}_{t} + \tilde{\mathfrak{R}}_{c} \right)/ 3 \eta_s \eta_d, \label{TotalRateEXP}\\
\eta_{c} &= \frac{8\, \tilde{\mathfrak{R}}_{c}/\tilde{\mathfrak{R}}_{t}}{\eta_s \eta_d (3+\tilde{\mathfrak{R}}_{c}/\tilde{\mathfrak{R}}_{t}) }. \label{EtaEXP1}
\end{align}

\subsection{Two SNSPD Configuration}
For the dual detector apparatus configuration, shown in Fig.~\ref{Expt}(b), component rates for the \(a\) and \(b\) photons can be measured separately and the subsequent analysis is more straightforward.  We use the total detector \(a\) rate \(\mathfrak{R}_{a}\) and the corresponding dark count rate \(D\) to determine the corrected arm \(a\) detection rate \(\tilde{\mathfrak{R}}_a = \mathfrak{R}_a - D\).  Similarly we use the total detector \(b\) rate \(\mathfrak{R}_{b}\) and the corresponding dark count rate \(D\) to determine the corrected arm \(b\) detection rate \(\tilde{\mathfrak{R}}_b = \mathfrak{R}_b - D\).  We use the detected coincident rate \(\mathfrak{R}_{c}\) and the corresponding coincident dark count rate \(D_{c}\) to determine the corrected detection rate for coincident photon pairs \(\tilde{\mathfrak{R}}_c = \mathfrak{R}_c - D_{c}\). Expressions for these corrected detection rates are as follows: 
\begin{align}
\tilde{\mathfrak{R}}_{a} = &R_{a} P(1|1), \label{ADetectedRate2}\\
\tilde{\mathfrak{R}}_{b} = &R_{b} P(1|1), \label{BDetectedRate2}\\
\tilde{\mathfrak{R}}_{c} = &R_{c} P(1|1) P(1|1). \label{PairDetectedRate2}
\end{align}
Again, we can invert these equations and substitute expressions from Tables \ref{2PhotonTable} and \ref{1PhotonTable} to find the true emission rates and \(\eta_{c}\) values in terms of \(\eta_s\), \(\eta_d\), and the measured detection rates:
\begin{align}
R_{a} &= \tilde{\mathfrak{R}}_{a}/\eta_s \eta_d, \label{ARateEXP}\\ 
R_{b} &= \tilde{\mathfrak{R}}_{b}/\eta_s \eta_d, \label{BRateEXP}\\ 
R_{c} &= \tilde{\mathfrak{R}}_{c}/\left(\eta_s \eta_d \right)^{2}, \label{CRateEXP2}\\ 
\eta_{c} &= \frac{1}{\eta_s \eta_d} \sqrt{\frac{\tilde{\mathfrak{R}}_{c}}{\tilde{\mathfrak{R}}_{a}} \frac{\tilde{\mathfrak{R}}_{c}}{\tilde{\mathfrak{R}}_{b}} }. \label{EtaEXP2}
\end{align}

\section{Experimental Results}
\begin{table*}
\setlength{\tabcolsep}{7pt}
\caption{Measured values of component system efficiencies, listed as percentage; detector dark counts, listed as kilocounts per second (kcps); and the corresponding uncertainties, are listed for the detection systems used.}
\label{EtaTable1}
\begin{ruledtabular}
\begin{tabular}{l | l l l l | l l | l l}
Configuration						& \(\eta_{d}\)		& \(\Delta \eta_{d}\)		& \(\Delta_{1}\)		& \(\Delta_{2}\)	& \(\eta_{s}\)	& \(\Delta \eta_{s}\)		& \(D\)													& \(\Delta D\)		\\ \hline
Single Detector					& \(56.3\%\) 			& \(0.24\%\)						& \(0.18\%\)				& \(0.16\%\)   		& \(91.5\%\)		& \(0.35\%\)						& \(1.4-2.4\, \mathrm{kcps}\)		& \(0.1\, \mathrm{kcps}\)\\
Detector \(a\) 					& \(67.9\%\) 			& \(0.61\%\)						& \(0.49\%\)				& \(0.26\%\)			& \(71.4\%\)		& \(0.37\%\)							& \(0.8\, \mathrm{kcps}\)				& \(0.1\, \mathrm{kcps}\) \\
Detector \(b\) 					& \(37.1\%\)			& \(0.29\%\)						& \(0.25\%\)				& \(0.14\%\)			& \(67.4\%\)		& \(0.33\%\)							& \(6.0\, \mathrm{kcps}\)				& \(1.0\, \mathrm{kcps}\)
\end{tabular}
\end{ruledtabular}
\end{table*}
To determine the photon emission rates and \(\eta_c\), we first characterized the system transmission efficiency \(\eta_s\), its uncertainty \(\Delta \eta_{s}\), the detector efficiency \(\eta_d\), its uncertainty \(\Delta \eta_{d}\), and the detector dark count rates \(D\) and its uncertainty \(\Delta D\).  These system characteristics, for both the single SNSPD configuration and the dual SNSPD configuration, are listed in Table \ref{EtaTable1}.

We determined \(\eta_s\) by sending the output of a stable, polarized CW 1560\,nm fiber laser through a polarization controller and coupling it to the optical system output fiber.  In this way we were able to reverse propagate the beam through the optical system; \(\eta_s\) was the ratio of the power before and after the optical system.  The transmission efficiency of the system---other than the fiber polarizing beamsplitter---was insensitive to the polarization of this input beam.  This characterization method ensured that the measured efficiency corresponded to the efficiency experienced by the collected modes.  The uncertainty \(\Delta \eta_{s}\) was due to the fluctuations in the power measurements.

We characterized the detector efficiency \(\eta_d\) by sending the same 1560\,nm laser through calibrated attenuators and a fiber based polarization controller and coupling it to the input fiber of the SNSPD apparatus; \(\eta_d\) was the ratio of photon detection rate---measured by the SNSPD, to the incident attenuated photon rate---known by measuring the unattenuated light with a power meter calibrated by NIST (National Institute of Standards and Technology) to \(0.38\%\) uncertainty.  The dominant contributions to the uncertainty \(\Delta \eta_{d}\) were the contributions due to the fluctuations in detected photons \(\Delta_{1}\), and the contributions due to the uncertainty of the calibrated power meter \(\Delta_{2}\), which are both listed in Table \ref{EtaTable1}.

We rotated the input polarization to find both the maximum and minimum polarization dependent detection efficiencies, which correspond to orthogonal input polarizations.  For the dual detector configuration, because both input polarizations were optimized separately, the relevant efficiency is the maximum efficiency.  However, for the single detector configuration, we operated with the polarization setting that maximized coincident photon detection rate---which is bounded from below by the lowest detection rate---suggesting that both of the orthogonally polarized SPDC photons were detected with the same, non-optimum, detection efficiency.  This occurs when the photons are polarized at \(\pm 45^{\circ}\) with respect to the nanowire pattern direction.  The relevant detector efficiency in this orientation for the single detector configuration is the average of the maximum and minimum polarization dependent efficiency \cite{dorenbos_superconducting_2008}. 

The output fiber of the optical system and the input fiber of the detector apparatus were then connected---either with a characterized and repeatable fiber coupler, or by fusion splicing the fiber ends and characterizing the splice loss (both methods were used). The detector dark count rates were determined by turning the pump laser off and recording the residual count rates in a darkened laboratory environment. This directly measured \(D\), as reported in Table \ref{EtaTable1}, as well as the contribution to \(D_c\) from coincidences between two dark counts, which were less than 1\,count per second.  The remaining portion of \(D_c\), resulting from coincidences between a single dark count and a single detected photon, for a given photon detection rate, is proportional to the product of \(D\), the total photon detection rate and the coincidence time-window of \(\Delta t =1\)\,ns.  For all of our data, this contribution to \(D_c\) was many orders of magnitude smaller than the measured coincident detection rate.  Because both contributions to \(D_c\) were far below measured coincident detection rate, we took \(D_c\) to be zero for our data. 

An important note is that detector \(b\) was located in a different room than that for the optical system and was connected using multiple single mode fiber patch cords strung together with four standard fiber couplers.  This optical fiber assembly was stable, however it was not optimized to maximize transmission.  Because we counted this lossy fiber assembly as part of the detector system itself, detector \(b\) exhibited a reduced detection efficiency.

\begin{figure}
\includegraphics[scale=0.85]{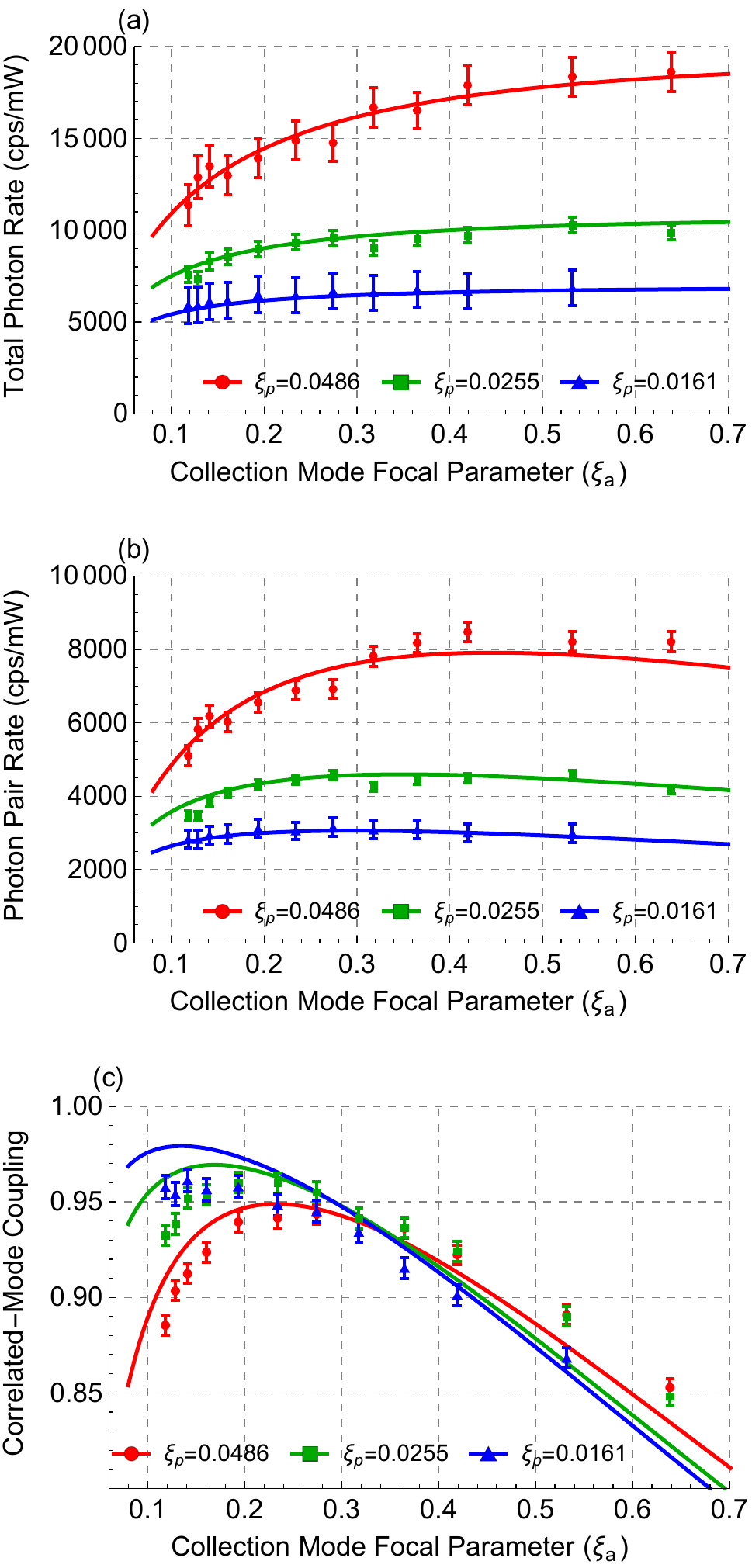}
\caption{\label{TotalPlot1}(color online) Single detector configuration data are shown as a function of collection mode focal parameter \(\xi_{a}\).  Plot (a) shows total photon emission rates in counts per second (cps) per mW of pump power, as a function of collection mode focal parameter \(\xi_{a}\), for three pump mode focal parameters \(\xi_{p}\).  Measured data, calculated from Eq.~(\ref{TotalRateEXP}) are shown as points and predicted behavior from Eq.~(\ref{TotalRateTH}) is shown as solid curves.   Plot (b) shows coincident photon pair emission rates in counts per second (cps) per mW of  pump power, as a function of collection mode focal parameter \(\xi_{a}\), for three pump mode focal parameters \(\xi_{p}\).  Measured data, calculated from Eq.~(\ref{CRateEXP1}) are shown as points and predicted behavior from Eq.~(\ref{CRateTH}) is shown as solid curves.  Plot (c) shows correlated-mode coupling efficiencies as a function of collection mode focal parameter \(\xi_{a}\), for three pump mode focal parameters \(\xi_{p}\).  Measured data, calculated from Eq.~(\ref{EtaEXP1}), are shown as points and predicted behavior from Eq.~(\ref{EtaTH}) is shown as solid curves.}
\end{figure}
For the single SNSPD configuration we measured the optimized total photon detection rate \(\mathfrak{R}_{t}\) and total coincident photon pair detection rate \(\mathfrak{R}_{c}\), over a range of collection mode focal parameters from 0.11 to 0.63 (corresponding to collection mode beam waists, at the center of the crystal, ranging from 47\,\(\mu\)m to 110\,\(\mu\)m)  for pump focal parameters of \(48.6\times10^{-3}\), \(25.5\times10^{-3}\), and \(16.1\times10^{-3}\) (corresponding to pump mode beam waists at the center of the crystal of 121\,\(\mu\)m, 167\,\(\mu\)m, and 209\,\(\mu\)m, respectively).  We then used Eqs.~(\ref{CRateTH})--(\ref{EtaTH}) to determine the corresponding emission rates and the correlated-mode coupling efficiencies. 
 
Both the measured total photon emission rates for the single SNSPD configuration, and the predicted total emission rates from Eq.~(\ref{TotalRateTH}) are displayed in Fig.~\ref{TotalPlot1}(a), for the various zoom lens configurations used.  Figure \ref{TotalPlot1}(b) displays both the measured photon pair emission rates as well as the predicted photon emission rates from Eq.~(\ref{CRateTH}).  The corresponding \(\eta_c\) measurements and theoretical predictions from Eq.~(\ref{EtaTH}) are displayed in Fig.~\ref{TotalPlot1}(c).  In these plots, measured data are shown as points and theoretical predictions are shown as solid curves.  

In this data, the effective nonlinearity \(d_{\mathrm{eff}}\) is treated as a free parameter.  The total photon rates and the coincident photon rates both rely quadratically on its magnitude, making this set of measurements an absolute characterization (i.e.\ independent of \(\eta_s\) or \(\eta_d\)) of the \(d_{\mathrm{eff}}\) for our PPKTP crystal, which we found to be \(d_{\mathrm{eff}} = 1.82\)\,pm/V.  We used this single value for our theoretical treatment, and although this is lower than the 2.3\,pm/V value which is typically reported, it is well within the range of values reported for absolute measurements \cite{Boulanger_Methodology_1998, Pack_Nonlinear_2004}.

We note that the plotted data have been corrected for detector dark counts, but not for other background counts, e.g.\ fluorescence counts from the crystal.  In this way we accurately describe the source's inherent capabilities, which could be realized by using more ideal detectors.  The error bars are found by looking at the range of data from many 1\,s integration windows.  The pump laser was not power stabilized and it fluctuated at both several minute time scales---which we accounted for by making repeated power measurements, and at hundreds of milliseconds time scales---which were the main source of noise for the data.

The data from this single SNSPD configuration agrees very well with the theoretical predictions with two small deviations. The first deviation is that for the largest collection mode focal parameters (smaller beam waists) we were able to achieve a few percent better than the predicted optimum \(\eta_{c}\) values.  We believe this to be the result of overestimating the effective mode focal parameter.  We note that the theoretical treatment assumes Gaussian collection modes, whereas single mode SMF-28 optical fibers support nearly---but not perfectly---Gaussian profile beams (see for example chapter 9 of \cite{SalehTeich_Photonics_2013}), providing some flexibility in establishing appropriate correspondence between the true mode width and the modeled Gaussian mode width.  We matched the modeled Gaussian mode beam waist to the measured true beam waist (the \(1/\mathrm{e}^2\) radius at the focus), which has an expected \(98.3\%\) mode overlap.  An alternative method of matching full widths at half maximum could have been employed.  Calculations indicate this method would give a Gaussian mode with a mode focal parameter that is smaller by about \(10\%\), a difference that would account for the data overshoot.  We are currently investigating non-Gaussian modes to further optimize fiber systems, as well as to incorporate optical waveguides whose mode profiles cannot be approximated by a Gaussian profile. 

The second deviation is that we were not able to achieve an \(\eta_{c}\) value above \(\sim\!\!96\%\)---even for the loosest pump focus of \(\xi_{p} = 16.1\times10^{-3}\), where we expect to achieve \(\eta_{c} = 98\%\).  This is a manifestation of the remaining background counts that we do not correct for.  We believe the reduction in the coupling efficiency is due to unpaired fluorescence photons from the PPKTP crystal. Previous measurements of a PPKTP waveguide at 1.3\,\(\mu\)m showed a fluorescence level of 2\% relative to SPDC (paired) photons within the phase-matching bandwidth \cite{Zhong_HighPerformance_2009}. The fluorescence amount is expected to be similar at the slightly longer wavelength that we operated in, suggesting the 2\% fluorescence accounts for the observed 2\% degradation in the correlated-mode coupling efficiency.

\begin{figure}
\includegraphics[scale=0.85]{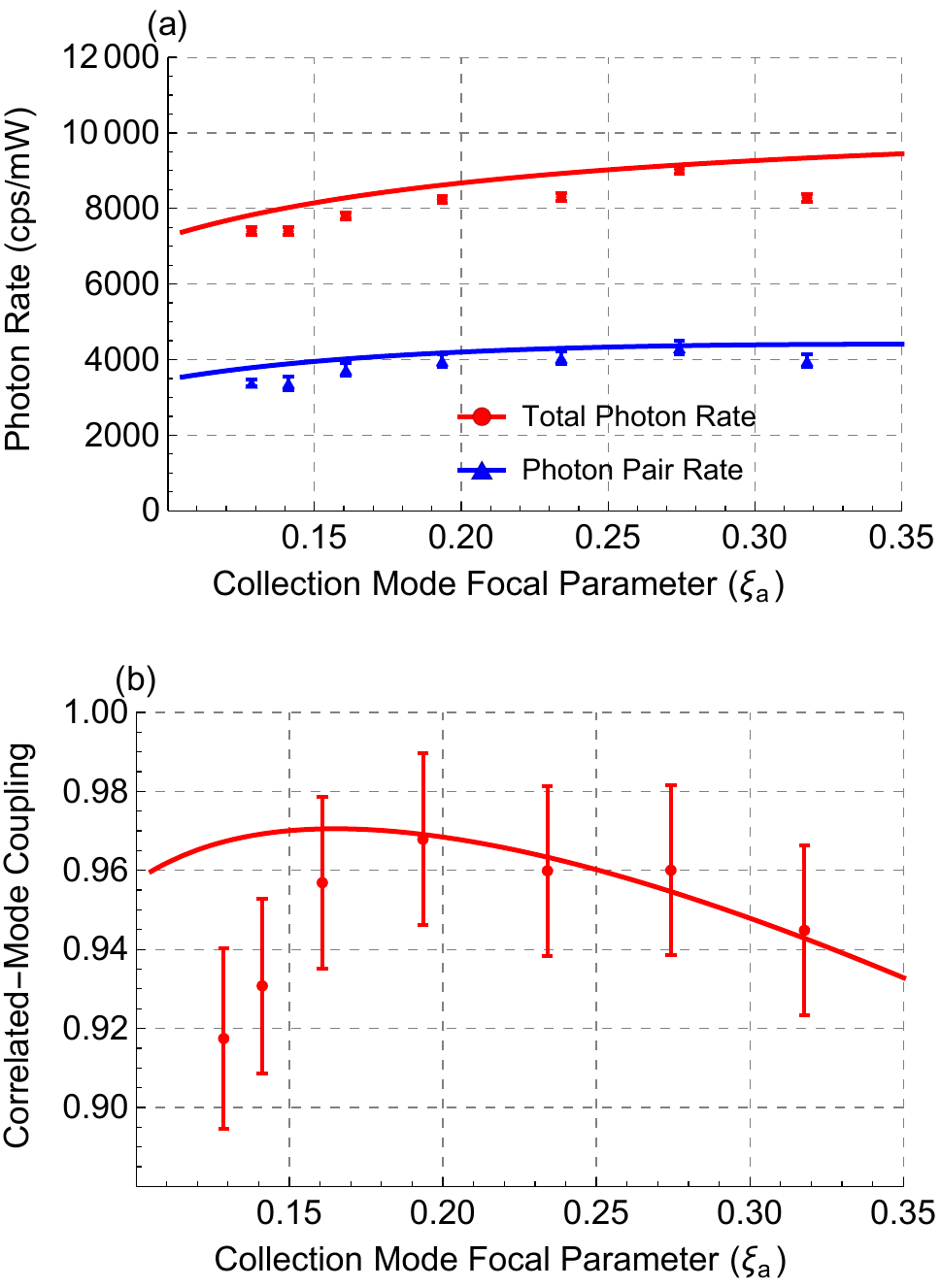}
\caption{\label{RatePlot2}(color online) Dual detector configuration data are shown as a function of collection mode focal parameter \(\xi_{a}\).  Plot (a) shows total photon rates and coincident photon pair rates in counts per second (cps) per mW of pump power, for pump mode focal parameter \(\xi_{p} = 0.0243\).  Measured data, calculated from Eqs.~(\ref{TotalRateEXP}) and (\ref{CRateEXP2}), are shown as points and predicted behaviors from Eqs.~(\ref{TotalRateTH}) and (\ref{CRateTH}) are shown as solid curves.   Plot (b) shows correlated-mode coupling efficiencies as a function of collection mode focal parameter \(\xi_{a}\), for pump mode focal parameter \(\xi_{p} = 0.0243\).  Measured data, calculated from Eq.~(\ref{EtaEXP2}), are shown as points and predicted behavior from Eq.~(\ref{EtaTH}) is shown as a solid curve.}
\end{figure}

To further verify the system performance we took additional data with the dual SNSPD configuration.  This configuration allowed us to independently measure the detection rates \(\mathfrak{R}_{a}\) and \(\mathfrak{R}_{b}\), for \(a\) and \(b\) photons, respectively, as well as the total coincident photon pair detection rate \(\mathfrak{R}_{c}\).  We made these measurements for a single pump beam focal parameter \(\xi_{p} = 24.3\times10^{-3}\) (corresponding to a pump mode beam waist at the center of the crystal of 169\,\(\mu\)m) over an abbreviated range of collected mode focal parameters.  We use Eqs.~(\ref{CRateTH})--(\ref{EtaTH}) to determine the emission rates shown in Fig.~\ref{RatePlot2}(a), and the correlated-mode coupling efficiencies shown in Fig.~\ref{RatePlot2}(b). In these plots, measured data are shown as points and theoretical predictions are shown as solid curves and the data have again been corrected for detector dark counts, but not for other background counts.  Additionally, the pump laser power again fluctuated at both several minute time scales---which we accounted for by making repeated power measurements, and at hundreds of milliseconds time scales---which was the main source of noise for the data.

The data from this configuration again agree very well with theoretical predictions and the high values of \(\eta_{c}\) were verified. Although it may seem that there is a systematic trend of reduced \(\eta_{c}\) values for the two smallest collection mode focal parameters, we believe this is simply statistical variation. We do not explicitly plot \(R_{a}\) or \(R_{b}\), but rather their sum, \(R_{t}\); however, for all points, these individual rates were within \(2\%\) of each other.  The background counts, however, were significantly different, with \(\sim\!\!1400\)\,cps background photon detection rate in arm \(a\) and \(\sim\!600\)\,cps background photon detection rate in arm \(b\).  This indicates that this background light is somewhat polarized. We achieved a peak \(\eta_{c}\) value of \(97 \pm 2\%\) at \(\xi_{a} = 0.19\), which had a corresponding one way mode coupling efficiency of \(P_{p}/P_{a} = 98\pm  1\%\).

\section{Discussion}
We have reported on a systematic investigation of the photon-pair spatial emission characteristics of a bulk nonlinear crystal coupled to single mode optical fibers.  We demonstrated a peak correlated-mode coupling efficiency of \(\eta_{c}=97\pm2\% \), which is significantly higher than the \(90\%\)--\(92\%\) peak efficiencies previously reported \cite{christensen_detection-loophole-free_2013, cunha_pereira_demonstrating_2013, giustina_bell_2013}.  Our results demonstrate both that the theoretical treatment of Bennink is a very good model of the system emission rates and correlated-mode coupling capabilities, and that high values for the correlated-mode coupling efficiency are ultimately achievable.  This indicates that high overall heralding efficiencies---highly desirable system characteristics for quantum information science applications---are achievable as well.  Moreover, because our system uses telecom band photons it can interface with existing telecom infrastructure, allowing it to be developed into a deployable optical system.

Perhaps a more instructive way to understand and compare correlated-mode coupling efficiencies is to consider it as an inherent source loss rather than transmission efficiency.  Indeed, this is a natural point of view for stringent loophole free Bell test measurements where we may have an overall loss budget of just \(17\%\) or 0.8\,dB \cite{dixon_max-ent_2014}.  The previously reported systems have \(8\%\)--\(10\%\) or about 0.4\,dB of inherent source loss \cite{christensen_detection-loophole-free_2013, cunha_pereira_demonstrating_2013, giustina_bell_2013}.  For this Bell test measurement then, the inherent source loss alone accounts for half of the overall loss budget.  Our demonstration reduces this inherent source loss by a factor of three to just over \(3\%\) or 0.14\,dB, leaving much more headroom to accommodate other system imperfections or detector inefficiencies.

The utility of this high correlated-mode coupling efficiency is not limited solely to stringent Bell tests.  An efficient heralded photon source is of use in applications that rely upon accurate knowledge of the source behavior.  Because of this broad range of applications, it is expected that these results and techniques to optimize \( \eta_{c} \) will be widely incorporated into future quantum optical research systems---a point that is illustrated by the fact that we are currently implementing these mode coupling techniques to develop sources to be used for polarization entangled photon generation, for quantum key distribution, for linear optical quantum computing, and for quantum illumination based sensing \cite{Guha_Gaussian_2009}.

\begin{acknowledgments}
This work is sponsored by the Assistant Secretary of Defense for Research and Engineering under Air Force Contract FA8721-05-C-0002. Opinions, interpretations, conclusions, and recommendations are those of the author and are not necessarily endorsed by the United States Government.
\end{acknowledgments}


\begin{thebibliography}{40}%
\makeatletter
\providecommand \@ifxundefined [1]{%
 \@ifx{#1\undefined}
}%
\providecommand \@ifnum [1]{%
 \ifnum #1\expandafter \@firstoftwo
 \else \expandafter \@secondoftwo
 \fi
}%
\providecommand \@ifx [1]{%
 \ifx #1\expandafter \@firstoftwo
 \else \expandafter \@secondoftwo
 \fi
}%
\providecommand \natexlab [1]{#1}%
\providecommand \enquote  [1]{``#1''}%
\providecommand \bibnamefont  [1]{#1}%
\providecommand \bibfnamefont [1]{#1}%
\providecommand \citenamefont [1]{#1}%
\providecommand \href@noop [0]{\@secondoftwo}%
\providecommand \href [0]{\begingroup \@sanitize@url \@href}%
\providecommand \@href[1]{\@@startlink{#1}\@@href}%
\providecommand \@@href[1]{\endgroup#1\@@endlink}%
\providecommand \@sanitize@url [0]{\catcode `\\12\catcode `\$12\catcode
  `\&12\catcode `\#12\catcode `\^12\catcode `\_12\catcode `\%12\relax}%
\providecommand \@@startlink[1]{}%
\providecommand \@@endlink[0]{}%
\providecommand \url  [0]{\begingroup\@sanitize@url \@url }%
\providecommand \@url [1]{\endgroup\@href {#1}{\urlprefix }}%
\providecommand \urlprefix  [0]{URL }%
\providecommand \Eprint [0]{\href }%
\providecommand \doibase [0]{http://dx.doi.org/}%
\providecommand \selectlanguage [0]{\@gobble}%
\providecommand \bibinfo  [0]{\@secondoftwo}%
\providecommand \bibfield  [0]{\@secondoftwo}%
\providecommand \translation [1]{[#1]}%
\providecommand \BibitemOpen [0]{}%
\providecommand \bibitemStop [0]{}%
\providecommand \bibitemNoStop [0]{.\EOS\space}%
\providecommand \EOS [0]{\spacefactor3000\relax}%
\providecommand \BibitemShut  [1]{\csname bibitem#1\endcsname}%
\let\auto@bib@innerbib\@empty
\bibitem [{\citenamefont {Knill}\ \emph {et~al.}(2001)\citenamefont {Knill},
  \citenamefont {Laflamme},\ and\ \citenamefont {Milburn}}]{knill_scheme_2001}%
  \BibitemOpen
  \bibfield  {author} {\bibinfo {author} {\bibfnamefont {E.}~\bibnamefont
  {Knill}}, \bibinfo {author} {\bibfnamefont {R.}~\bibnamefont {Laflamme}}, \
  and\ \bibinfo {author} {\bibfnamefont {G.~J.}\ \bibnamefont {Milburn}},\
  }\href {\doibase 10.1038/35051009} {\bibfield  {journal} {\bibinfo  {journal}
  {Nature}\ }\textbf {\bibinfo {volume} {409}},\ \bibinfo {pages} {46}
  (\bibinfo {year} {2001})}\BibitemShut {NoStop}%
\bibitem [{\citenamefont {Kok}\ \emph {et~al.}(2007)\citenamefont {Kok},
  \citenamefont {Munro}, \citenamefont {Nemoto}, \citenamefont {Ralph},
  \citenamefont {Dowling},\ and\ \citenamefont {Milburn}}]{kok_linear_2007}%
  \BibitemOpen
  \bibfield  {author} {\bibinfo {author} {\bibfnamefont {P.}~\bibnamefont
  {Kok}}, \bibinfo {author} {\bibfnamefont {W.~J.}\ \bibnamefont {Munro}},
  \bibinfo {author} {\bibfnamefont {K.}~\bibnamefont {Nemoto}}, \bibinfo
  {author} {\bibfnamefont {T.~C.}\ \bibnamefont {Ralph}}, \bibinfo {author}
  {\bibfnamefont {J.~P.}\ \bibnamefont {Dowling}}, \ and\ \bibinfo {author}
  {\bibfnamefont {G.~J.}\ \bibnamefont {Milburn}},\ }\href {\doibase
  10.1103/RevModPhys.79.135} {\bibfield  {journal} {\bibinfo  {journal} {Rev.\
  Mod.\ Phys.}\ }\textbf {\bibinfo {volume} {79}},\ \bibinfo {pages} {135}
  (\bibinfo {year} {2007})}\BibitemShut {NoStop}%
\bibitem [{\citenamefont {Aaronson}\ and\ \citenamefont
  {Arkhipov}(2011)}]{aaronson_computational_2011}%
  \BibitemOpen
  \bibfield  {author} {\bibinfo {author} {\bibfnamefont {S.}~\bibnamefont
  {Aaronson}}\ and\ \bibinfo {author} {\bibfnamefont {A.}~\bibnamefont
  {Arkhipov}},\ }in\ \href {\doibase 10.1145/1993636.1993682} {\emph {\bibinfo
  {booktitle} {Proc.\ of the Forty-third Annual {ACM} Symposium on Theory of
  Computing}}},\ \bibinfo {series and number} {{STOC} '11}\ (\bibinfo
  {publisher} {{ACM}},\ \bibinfo {year} {2011})\ p.\ \bibinfo {pages}
  {333–342}\BibitemShut {NoStop}%
\bibitem [{\citenamefont {Ralph}(2013)}]{ralph_quantum_2013}%
  \BibitemOpen
  \bibfield  {author} {\bibinfo {author} {\bibfnamefont {T.~C.}\ \bibnamefont
  {Ralph}},\ }\href {\doibase 10.1038/nphoton.2013.175} {\bibfield  {journal}
  {\bibinfo  {journal} {Nature Photon.}\ }\textbf {\bibinfo {volume} {7}},\
  \bibinfo {pages} {514} (\bibinfo {year} {2013})}\BibitemShut {NoStop}%
\bibitem [{\citenamefont {Bennett}\ and\ \citenamefont
  {Brassard}(1984)}]{bennett_quantum_1984}%
  \BibitemOpen
  \bibfield  {author} {\bibinfo {author} {\bibfnamefont {C.~H.}\ \bibnamefont
  {Bennett}}\ and\ \bibinfo {author} {\bibfnamefont {G.}~\bibnamefont
  {Brassard}},\ }in\ \href@noop {} {\emph {\bibinfo {booktitle} {Proc.\ IEEE
  Int.\ Conf.\ on Computers, Systems \& Signal Processing, Bangalore, India}}}\
  (\bibinfo  {publisher} {{IEEE, New York}},\ \bibinfo {year}
  {1984})\BibitemShut {NoStop}%
\bibitem [{\citenamefont {Ekert}(1991)}]{ekert_quantum_1991}%
  \BibitemOpen
  \bibfield  {author} {\bibinfo {author} {\bibfnamefont {A.~K.}\ \bibnamefont
  {Ekert}},\ }\href {\doibase 10.1103/PhysRevLett.67.661} {\bibfield  {journal}
  {\bibinfo  {journal} {Phys.\ Rev.\ Lett.}\ }\textbf {\bibinfo {volume}
  {67}},\ \bibinfo {pages} {661} (\bibinfo {year} {1991})}\BibitemShut
  {NoStop}%
\bibitem [{\citenamefont {Hughes}\ and\ \citenamefont
  {Nordholt}(2011)}]{hughes_refining_2011}%
  \BibitemOpen
  \bibfield  {author} {\bibinfo {author} {\bibfnamefont {R.}~\bibnamefont
  {Hughes}}\ and\ \bibinfo {author} {\bibfnamefont {J.}~\bibnamefont
  {Nordholt}},\ }\href {\doibase 10.1126/science.1208527} {\bibfield  {journal}
  {\bibinfo  {journal} {Science}\ }\textbf {\bibinfo {volume} {333}},\ \bibinfo
  {pages} {1584} (\bibinfo {year} {2011})}\BibitemShut {NoStop}%
\bibitem [{\citenamefont {Weedbrook}\ \emph {et~al.}(2012)\citenamefont
  {Weedbrook}, \citenamefont {Pirandola}, \citenamefont
  {Garc\'{i}a-Patr\'{o}n}, \citenamefont {Cerf}, \citenamefont {Ralph},
  \citenamefont {Shapiro},\ and\ \citenamefont
  {Lloyd}}]{weedbrook_gaussian_2012}%
  \BibitemOpen
  \bibfield  {author} {\bibinfo {author} {\bibfnamefont {C.}~\bibnamefont
  {Weedbrook}}, \bibinfo {author} {\bibfnamefont {S.}~\bibnamefont
  {Pirandola}}, \bibinfo {author} {\bibfnamefont {R.}~\bibnamefont
  {Garc\'{i}a-Patr\'{o}n}}, \bibinfo {author} {\bibfnamefont {N.~J.}\
  \bibnamefont {Cerf}}, \bibinfo {author} {\bibfnamefont {T.~C.}\ \bibnamefont
  {Ralph}}, \bibinfo {author} {\bibfnamefont {J.~H.}\ \bibnamefont {Shapiro}},
  \ and\ \bibinfo {author} {\bibfnamefont {S.}~\bibnamefont {Lloyd}},\ }\href
  {\doibase 10.1103/RevModPhys.84.621} {\bibfield  {journal} {\bibinfo
  {journal} {Rev.\ Mod.\ Phys.}\ }\textbf {\bibinfo {volume} {84}},\ \bibinfo
  {pages} {621} (\bibinfo {year} {2012})}\BibitemShut {NoStop}%
\bibitem [{\citenamefont {Sakurai}(1993)}]{Sakurai_Modern_1993}%
  \BibitemOpen
  \bibfield  {author} {\bibinfo {author} {\bibfnamefont {J.~J.}\ \bibnamefont
  {Sakurai}},\ }\href@noop {} {\emph {\bibinfo {title} {{Modern Quantum
  Mechanics (Revised Edition)}}}},\ \bibinfo {edition} {1st}\ ed.\ (\bibinfo
  {publisher} {Addison Wesley},\ \bibinfo {year} {1993})\BibitemShut {NoStop}%
\bibitem [{\citenamefont {Bell}(1964)}]{Bell_EPR_1964}%
  \BibitemOpen
  \bibfield  {author} {\bibinfo {author} {\bibfnamefont {J.~S.}\ \bibnamefont
  {Bell}},\ }\href@noop {} {\bibfield  {journal} {\bibinfo  {journal}
  {Physics}\ }\textbf {\bibinfo {volume} {1}},\ \bibinfo {pages} {195}
  (\bibinfo {year} {1964})}\BibitemShut {NoStop}%
\bibitem [{\citenamefont {Brunner}\ \emph {et~al.}(2014)\citenamefont
  {Brunner}, \citenamefont {Cavalcanti}, \citenamefont {Pironio}, \citenamefont
  {Scarani},\ and\ \citenamefont {Wehner}}]{Brunner_Bell_2014}%
  \BibitemOpen
  \bibfield  {author} {\bibinfo {author} {\bibfnamefont {N.}~\bibnamefont
  {Brunner}}, \bibinfo {author} {\bibfnamefont {D.}~\bibnamefont {Cavalcanti}},
  \bibinfo {author} {\bibfnamefont {S.}~\bibnamefont {Pironio}}, \bibinfo
  {author} {\bibfnamefont {V.}~\bibnamefont {Scarani}}, \ and\ \bibinfo
  {author} {\bibfnamefont {S.}~\bibnamefont {Wehner}},\ }\href {\doibase
  10.1103/RevModPhys.86.419} {\bibfield  {journal} {\bibinfo  {journal} {Rev.\
  Mod.\ Phys.}\ }\textbf {\bibinfo {volume} {86}},\ \bibinfo {pages} {419}
  (\bibinfo {year} {2014})}\BibitemShut {NoStop}%
\bibitem [{\citenamefont {Wittmann}\ \emph {et~al.}(2012)\citenamefont
  {Wittmann}, \citenamefont {Ramelow}, \citenamefont {Steinlechner},
  \citenamefont {Langford}, \citenamefont {Brunner}, \citenamefont {Wiseman},
  \citenamefont {Ursin},\ and\ \citenamefont
  {Zeilinger}}]{wittmann_loophole-free_2012}%
  \BibitemOpen
  \bibfield  {author} {\bibinfo {author} {\bibfnamefont {B.}~\bibnamefont
  {Wittmann}}, \bibinfo {author} {\bibfnamefont {S.}~\bibnamefont {Ramelow}},
  \bibinfo {author} {\bibfnamefont {F.}~\bibnamefont {Steinlechner}}, \bibinfo
  {author} {\bibfnamefont {N.~K.}\ \bibnamefont {Langford}}, \bibinfo {author}
  {\bibfnamefont {N.}~\bibnamefont {Brunner}}, \bibinfo {author} {\bibfnamefont
  {H.~M.}\ \bibnamefont {Wiseman}}, \bibinfo {author} {\bibfnamefont
  {R.}~\bibnamefont {Ursin}}, \ and\ \bibinfo {author} {\bibfnamefont
  {A.}~\bibnamefont {Zeilinger}},\ }\href {\doibase
  10.1088/1367-2630/14/5/053030} {\bibfield  {journal} {\bibinfo  {journal}
  {New J.\ Phys.}\ }\textbf {\bibinfo {volume} {14}},\ \bibinfo {pages}
  {053030} (\bibinfo {year} {2012})}\BibitemShut {NoStop}%
\bibitem [{\citenamefont {Christensen}\ \emph {et~al.}(2013)\citenamefont
  {Christensen}, \citenamefont {McCusker}, \citenamefont {Altepeter},
  \citenamefont {Calkins}, \citenamefont {Gerrits}, \citenamefont {Lita},
  \citenamefont {Miller}, \citenamefont {Shalm}, \citenamefont {Zhang},
  \citenamefont {Nam}, \citenamefont {Brunner}, \citenamefont {Lim},
  \citenamefont {Gisin},\ and\ \citenamefont
  {Kwiat}}]{christensen_detection-loophole-free_2013}%
  \BibitemOpen
  \bibfield  {author} {\bibinfo {author} {\bibfnamefont {B.~G.}\ \bibnamefont
  {Christensen}}, \bibinfo {author} {\bibfnamefont {K.~T.}\ \bibnamefont
  {McCusker}}, \bibinfo {author} {\bibfnamefont {J.~B.}\ \bibnamefont
  {Altepeter}}, \bibinfo {author} {\bibfnamefont {B.}~\bibnamefont {Calkins}},
  \bibinfo {author} {\bibfnamefont {T.}~\bibnamefont {Gerrits}}, \bibinfo
  {author} {\bibfnamefont {A.~E.}\ \bibnamefont {Lita}}, \bibinfo {author}
  {\bibfnamefont {A.}~\bibnamefont {Miller}}, \bibinfo {author} {\bibfnamefont
  {L.~K.}\ \bibnamefont {Shalm}}, \bibinfo {author} {\bibfnamefont
  {Y.}~\bibnamefont {Zhang}}, \bibinfo {author} {\bibfnamefont {S.~W.}\
  \bibnamefont {Nam}}, \bibinfo {author} {\bibfnamefont {N.}~\bibnamefont
  {Brunner}}, \bibinfo {author} {\bibfnamefont {C.~C.~W.}\ \bibnamefont {Lim}},
  \bibinfo {author} {\bibfnamefont {N.}~\bibnamefont {Gisin}}, \ and\ \bibinfo
  {author} {\bibfnamefont {P.~G.}\ \bibnamefont {Kwiat}},\ }\href {\doibase
  10.1103/PhysRevLett.111.130406} {\bibfield  {journal} {\bibinfo  {journal}
  {Phys.\ Rev.\ Lett.}\ }\textbf {\bibinfo {volume} {111}},\ \bibinfo {pages}
  {130406} (\bibinfo {year} {2013})}\BibitemShut {NoStop}%
\bibitem [{\citenamefont {Giustina}\ \emph {et~al.}(2013)\citenamefont
  {Giustina}, \citenamefont {Mech}, \citenamefont {Ramelow}, \citenamefont
  {Wittmann}, \citenamefont {Kofler}, \citenamefont {Beyer}, \citenamefont
  {Lita}, \citenamefont {Calkins}, \citenamefont {Gerrits}, \citenamefont
  {Nam}, \citenamefont {Ursin},\ and\ \citenamefont
  {Zeilinger}}]{giustina_bell_2013}%
  \BibitemOpen
  \bibfield  {author} {\bibinfo {author} {\bibfnamefont {M.}~\bibnamefont
  {Giustina}}, \bibinfo {author} {\bibfnamefont {A.}~\bibnamefont {Mech}},
  \bibinfo {author} {\bibfnamefont {S.}~\bibnamefont {Ramelow}}, \bibinfo
  {author} {\bibfnamefont {B.}~\bibnamefont {Wittmann}}, \bibinfo {author}
  {\bibfnamefont {J.}~\bibnamefont {Kofler}}, \bibinfo {author} {\bibfnamefont
  {J.}~\bibnamefont {Beyer}}, \bibinfo {author} {\bibfnamefont
  {A.}~\bibnamefont {Lita}}, \bibinfo {author} {\bibfnamefont {B.}~\bibnamefont
  {Calkins}}, \bibinfo {author} {\bibfnamefont {T.}~\bibnamefont {Gerrits}},
  \bibinfo {author} {\bibfnamefont {S.~W.}\ \bibnamefont {Nam}}, \bibinfo
  {author} {\bibfnamefont {R.}~\bibnamefont {Ursin}}, \ and\ \bibinfo {author}
  {\bibfnamefont {A.}~\bibnamefont {Zeilinger}},\ }\href {\doibase
  10.1038/nature12012} {\bibfield  {journal} {\bibinfo  {journal} {Nature}\
  }\textbf {\bibinfo {volume} {497}},\ \bibinfo {pages} {227} (\bibinfo {year}
  {2013})}\BibitemShut {NoStop}%
\bibitem [{\citenamefont {Eberhard}(1993)}]{eberhard_background_1993}%
  \BibitemOpen
  \bibfield  {author} {\bibinfo {author} {\bibfnamefont {P.~H.}\ \bibnamefont
  {Eberhard}},\ }\href {\doibase 10.1103/PhysRevA.47.R747} {\bibfield
  {journal} {\bibinfo  {journal} {Phys.\ Rev.\ A}\ }\textbf {\bibinfo {volume}
  {47}},\ \bibinfo {pages} {R747} (\bibinfo {year} {1993})}\BibitemShut
  {NoStop}%
\bibitem [{\citenamefont {Weihs}\ \emph {et~al.}(1998)\citenamefont {Weihs},
  \citenamefont {Jennewein}, \citenamefont {Simon}, \citenamefont
  {Weinfurter},\ and\ \citenamefont {Zeilinger}}]{weihs_violation_1998}%
  \BibitemOpen
  \bibfield  {author} {\bibinfo {author} {\bibfnamefont {G.}~\bibnamefont
  {Weihs}}, \bibinfo {author} {\bibfnamefont {T.}~\bibnamefont {Jennewein}},
  \bibinfo {author} {\bibfnamefont {C.}~\bibnamefont {Simon}}, \bibinfo
  {author} {\bibfnamefont {H.}~\bibnamefont {Weinfurter}}, \ and\ \bibinfo
  {author} {\bibfnamefont {A.}~\bibnamefont {Zeilinger}},\ }\href {\doibase
  10.1103/PhysRevLett.81.5039} {\bibfield  {journal} {\bibinfo  {journal}
  {Phys.\ Rev.\ Lett.}\ }\textbf {\bibinfo {volume} {81}},\ \bibinfo {pages}
  {5039} (\bibinfo {year} {1998})}\BibitemShut {NoStop}%
\bibitem [{\citenamefont {Rosenberg}\ \emph {et~al.}(2013)\citenamefont
  {Rosenberg}, \citenamefont {Kerman}, \citenamefont {Molnar},\ and\
  \citenamefont {Dauler}}]{rosenberg_high-speed_2013}%
  \BibitemOpen
  \bibfield  {author} {\bibinfo {author} {\bibfnamefont {D.}~\bibnamefont
  {Rosenberg}}, \bibinfo {author} {\bibfnamefont {A.~J.}\ \bibnamefont
  {Kerman}}, \bibinfo {author} {\bibfnamefont {R.~J.}\ \bibnamefont {Molnar}},
  \ and\ \bibinfo {author} {\bibfnamefont {E.~A.}\ \bibnamefont {Dauler}},\
  }\href {\doibase 10.1364/OE.21.001440} {\bibfield  {journal} {\bibinfo
  {journal} {Opt.\ Express}\ }\textbf {\bibinfo {volume} {21}},\ \bibinfo
  {pages} {1440} (\bibinfo {year} {2013})}\BibitemShut {NoStop}%
\bibitem [{\citenamefont {Marsili}\ \emph {et~al.}(2013)\citenamefont
  {Marsili}, \citenamefont {Verma}, \citenamefont {Stern}, \citenamefont
  {Harrington}, \citenamefont {Lita}, \citenamefont {Gerrits}, \citenamefont
  {Vayshenker}, \citenamefont {Baek}, \citenamefont {Shaw}, \citenamefont
  {Mirin},\ and\ \citenamefont {Nam}}]{marsili_detecting_2013}%
  \BibitemOpen
  \bibfield  {author} {\bibinfo {author} {\bibfnamefont {F.}~\bibnamefont
  {Marsili}}, \bibinfo {author} {\bibfnamefont {V.~B.}\ \bibnamefont {Verma}},
  \bibinfo {author} {\bibfnamefont {J.~A.}\ \bibnamefont {Stern}}, \bibinfo
  {author} {\bibfnamefont {S.}~\bibnamefont {Harrington}}, \bibinfo {author}
  {\bibfnamefont {A.~E.}\ \bibnamefont {Lita}}, \bibinfo {author}
  {\bibfnamefont {T.}~\bibnamefont {Gerrits}}, \bibinfo {author} {\bibfnamefont
  {I.}~\bibnamefont {Vayshenker}}, \bibinfo {author} {\bibfnamefont
  {B.}~\bibnamefont {Baek}}, \bibinfo {author} {\bibfnamefont {M.~D.}\
  \bibnamefont {Shaw}}, \bibinfo {author} {\bibfnamefont {R.~P.}\ \bibnamefont
  {Mirin}}, \ and\ \bibinfo {author} {\bibfnamefont {S.~W.}\ \bibnamefont
  {Nam}},\ }\href {\doibase 10.1038/nphoton.2013.13} {\bibfield  {journal}
  {\bibinfo  {journal} {Nature Photon.}\ }\textbf {\bibinfo {volume} {7}},\
  \bibinfo {pages} {210} (\bibinfo {year} {2013})}\BibitemShut {NoStop}%
\bibitem [{\citenamefont {Miki}\ \emph {et~al.}(2013)\citenamefont {Miki},
  \citenamefont {Yamashita}, \citenamefont {Terai},\ and\ \citenamefont
  {Wang}}]{miki_high_2013}%
  \BibitemOpen
  \bibfield  {author} {\bibinfo {author} {\bibfnamefont {S.}~\bibnamefont
  {Miki}}, \bibinfo {author} {\bibfnamefont {T.}~\bibnamefont {Yamashita}},
  \bibinfo {author} {\bibfnamefont {H.}~\bibnamefont {Terai}}, \ and\ \bibinfo
  {author} {\bibfnamefont {Z.}~\bibnamefont {Wang}},\ }\href {\doibase
  10.1364/OE.21.010208} {\bibfield  {journal} {\bibinfo  {journal} {Opt.\
  Express}\ }\textbf {\bibinfo {volume} {21}},\ \bibinfo {pages} {10208}
  (\bibinfo {year} {2013})}\BibitemShut {NoStop}%
\bibitem [{\citenamefont {Ljunggren}\ and\ \citenamefont
  {Tengner}(2005)}]{ljunggren_optimal_2005}%
  \BibitemOpen
  \bibfield  {author} {\bibinfo {author} {\bibfnamefont {D.}~\bibnamefont
  {Ljunggren}}\ and\ \bibinfo {author} {\bibfnamefont {M.}~\bibnamefont
  {Tengner}},\ }\href {\doibase 10.1103/PhysRevA.72.062301} {\bibfield
  {journal} {\bibinfo  {journal} {Phys.\ Rev.\ A}\ }\textbf {\bibinfo {volume}
  {72}},\ \bibinfo {pages} {062301} (\bibinfo {year} {2005})}\BibitemShut
  {NoStop}%
\bibitem [{\citenamefont {Mosley}\ \emph {et~al.}(2008)\citenamefont {Mosley},
  \citenamefont {Lundeen}, \citenamefont {Smith}, \citenamefont {Wasylczyk},
  \citenamefont {U'Ren}, \citenamefont {Silberhorn},\ and\ \citenamefont
  {Walmsley}}]{mosley_heralded_2008}%
  \BibitemOpen
  \bibfield  {author} {\bibinfo {author} {\bibfnamefont {P.~J.}\ \bibnamefont
  {Mosley}}, \bibinfo {author} {\bibfnamefont {J.~S.}\ \bibnamefont {Lundeen}},
  \bibinfo {author} {\bibfnamefont {B.~J.}\ \bibnamefont {Smith}}, \bibinfo
  {author} {\bibfnamefont {P.}~\bibnamefont {Wasylczyk}}, \bibinfo {author}
  {\bibfnamefont {A.~B.}\ \bibnamefont {U'Ren}}, \bibinfo {author}
  {\bibfnamefont {C.}~\bibnamefont {Silberhorn}}, \ and\ \bibinfo {author}
  {\bibfnamefont {I.~A.}\ \bibnamefont {Walmsley}},\ }\href {\doibase
  10.1103/PhysRevLett.100.133601} {\bibfield  {journal} {\bibinfo  {journal}
  {Phys.\ Rev.\ Lett.}\ }\textbf {\bibinfo {volume} {100}},\ \bibinfo {pages}
  {133601} (\bibinfo {year} {2008})}\BibitemShut {NoStop}%
\bibitem [{\citenamefont {Bennink}(2010)}]{bennink_optimal_2010}%
  \BibitemOpen
  \bibfield  {author} {\bibinfo {author} {\bibfnamefont {R.~S.}\ \bibnamefont
  {Bennink}},\ }\href {\doibase 10.1103/PhysRevA.81.053805} {\bibfield
  {journal} {\bibinfo  {journal} {Phys.\ Rev.\ A}\ }\textbf {\bibinfo {volume}
  {81}},\ \bibinfo {pages} {053805} (\bibinfo {year} {2010})}\BibitemShut
  {NoStop}%
\bibitem [{\citenamefont {Cunha~Pereira}\ \emph {et~al.}(2013)\citenamefont
  {Cunha~Pereira}, \citenamefont {Becerra}, \citenamefont {Glebov},
  \citenamefont {Fan}, \citenamefont {Nam},\ and\ \citenamefont
  {Migdall}}]{cunha_pereira_demonstrating_2013}%
  \BibitemOpen
  \bibfield  {author} {\bibinfo {author} {\bibfnamefont {M.~D.}\ \bibnamefont
  {Cunha~Pereira}}, \bibinfo {author} {\bibfnamefont {F.~E.}\ \bibnamefont
  {Becerra}}, \bibinfo {author} {\bibfnamefont {B.~L.}\ \bibnamefont {Glebov}},
  \bibinfo {author} {\bibfnamefont {J.}~\bibnamefont {Fan}}, \bibinfo {author}
  {\bibfnamefont {S.~W.}\ \bibnamefont {Nam}}, \ and\ \bibinfo {author}
  {\bibfnamefont {A.}~\bibnamefont {Migdall}},\ }\href {\doibase
  10.1364/OL.38.001609} {\bibfield  {journal} {\bibinfo  {journal} {Opt.\
  Lett.}\ }\textbf {\bibinfo {volume} {38}},\ \bibinfo {pages} {1609} (\bibinfo
  {year} {2013})}\BibitemShut {NoStop}%
\bibitem [{\citenamefont {Guerreiro}\ \emph {et~al.}(2013)\citenamefont
  {Guerreiro}, \citenamefont {Martin}, \citenamefont {Sanguinetti},
  \citenamefont {Bruno}, \citenamefont {Zbinden},\ and\ \citenamefont
  {Thew}}]{Guerreiro_HighEfficiency_2013}%
  \BibitemOpen
  \bibfield  {author} {\bibinfo {author} {\bibfnamefont {T.}~\bibnamefont
  {Guerreiro}}, \bibinfo {author} {\bibfnamefont {A.}~\bibnamefont {Martin}},
  \bibinfo {author} {\bibfnamefont {B.}~\bibnamefont {Sanguinetti}}, \bibinfo
  {author} {\bibfnamefont {N.}~\bibnamefont {Bruno}}, \bibinfo {author}
  {\bibfnamefont {H.}~\bibnamefont {Zbinden}}, \ and\ \bibinfo {author}
  {\bibfnamefont {R.~T.}\ \bibnamefont {Thew}},\ }\href {\doibase
  10.1364/OE.21.027641} {\bibfield  {journal} {\bibinfo  {journal} {Opt.
  Express}\ }\textbf {\bibinfo {volume} {21}},\ \bibinfo {pages} {27641}
  (\bibinfo {year} {2013})}\BibitemShut {NoStop}%
\bibitem [{\citenamefont {Shields}(2007)}]{shields_semiconductor_2007}%
  \BibitemOpen
  \bibfield  {author} {\bibinfo {author} {\bibfnamefont {A.~J.}\ \bibnamefont
  {Shields}},\ }\href {\doibase 10.1038/nphoton.2007.46} {\bibfield  {journal}
  {\bibinfo  {journal} {Nature Photon.}\ }\textbf {\bibinfo {volume} {1}},\
  \bibinfo {pages} {215} (\bibinfo {year} {2007})}\BibitemShut {NoStop}%
\bibitem [{\citenamefont {Pomarico}\ \emph {et~al.}(2012)\citenamefont
  {Pomarico}, \citenamefont {Sanguinetti}, \citenamefont {Guerreiro},
  \citenamefont {Thew},\ and\ \citenamefont {Zbinden}}]{pomarico_mhz_2012}%
  \BibitemOpen
  \bibfield  {author} {\bibinfo {author} {\bibfnamefont {E.}~\bibnamefont
  {Pomarico}}, \bibinfo {author} {\bibfnamefont {B.}~\bibnamefont
  {Sanguinetti}}, \bibinfo {author} {\bibfnamefont {T.}~\bibnamefont
  {Guerreiro}}, \bibinfo {author} {\bibfnamefont {R.}~\bibnamefont {Thew}}, \
  and\ \bibinfo {author} {\bibfnamefont {H.}~\bibnamefont {Zbinden}},\ }\href
  {\doibase 10.1364/OE.20.023846} {\bibfield  {journal} {\bibinfo  {journal}
  {Opt.\ Express}\ }\textbf {\bibinfo {volume} {20}},\ \bibinfo {pages} {23846}
  (\bibinfo {year} {2012})}\BibitemShut {NoStop}%
\bibitem [{\citenamefont {Mandel}\ and\ \citenamefont
  {Wolf}(1995)}]{mandel_optical_1995}%
  \BibitemOpen
  \bibfield  {author} {\bibinfo {author} {\bibfnamefont {L.}~\bibnamefont
  {Mandel}}\ and\ \bibinfo {author} {\bibfnamefont {E.}~\bibnamefont {Wolf}},\
  }\href@noop {} {\emph {\bibinfo {title} {Optical Coherence and Quantum
  Optics}}}\ (\bibinfo  {publisher} {Cambridge University Press},\ \bibinfo
  {year} {1995})\BibitemShut {NoStop}%
\bibitem [{\citenamefont {Hong}\ and\ \citenamefont
  {Mandel}(1986)}]{hong_experimental_1986}%
  \BibitemOpen
  \bibfield  {author} {\bibinfo {author} {\bibfnamefont {C.~K.}\ \bibnamefont
  {Hong}}\ and\ \bibinfo {author} {\bibfnamefont {L.}~\bibnamefont {Mandel}},\
  }\href {\doibase 10.1103/PhysRevLett.56.58} {\bibfield  {journal} {\bibinfo
  {journal} {Phys.\ Rev.\ Lett.}\ }\textbf {\bibinfo {volume} {56}},\ \bibinfo
  {pages} {58} (\bibinfo {year} {1986})}\BibitemShut {NoStop}%
\bibitem [{Dix()}]{DixonBennink_emails_2014}%
  \BibitemOpen
  \href@noop {} {}\bibinfo {note} {The derivation of these equations accounts
  for the explicit inclusion of the material index of refraction in the
  denominators of the quantized field definitions, which only appeared
  implicitly in the original derivation}\BibitemShut {NoStop}%
\bibitem [{\citenamefont {Boyd}(2008)}]{Boyd_nonlinear_2008}%
  \BibitemOpen
  \bibfield  {author} {\bibinfo {author} {\bibfnamefont {R.~W.}\ \bibnamefont
  {Boyd}},\ }\href@noop {} {\emph {\bibinfo {title} {Nonlinear Optics, Third
  Edition}}},\ \bibinfo {edition} {3rd}\ ed.\ (\bibinfo  {publisher} {Academic
  Press},\ \bibinfo {year} {2008})\BibitemShut {NoStop}%
\bibitem [{\citenamefont {Boyd}\ and\ \citenamefont
  {Kleinman}(1968)}]{boyd_parametric_1968}%
  \BibitemOpen
  \bibfield  {author} {\bibinfo {author} {\bibfnamefont {G.~D.}\ \bibnamefont
  {Boyd}}\ and\ \bibinfo {author} {\bibfnamefont {D.~A.}\ \bibnamefont
  {Kleinman}},\ }\href {\doibase 10.1063/1.1656831} {\bibfield  {journal}
  {\bibinfo  {journal} {J.\ Appl.\ Phys.}\ }\textbf {\bibinfo {volume} {39}},\
  \bibinfo {pages} {3597} (\bibinfo {year} {1968})}\BibitemShut {NoStop}%
\bibitem [{\citenamefont {Fradkin}\ \emph {et~al.}(1999)\citenamefont
  {Fradkin}, \citenamefont {Arie}, \citenamefont {Skliar},\ and\ \citenamefont
  {Rosenman}}]{Fradkin_Tunable_1999}%
  \BibitemOpen
  \bibfield  {author} {\bibinfo {author} {\bibfnamefont {K.}~\bibnamefont
  {Fradkin}}, \bibinfo {author} {\bibfnamefont {A.}~\bibnamefont {Arie}},
  \bibinfo {author} {\bibfnamefont {A.}~\bibnamefont {Skliar}}, \ and\ \bibinfo
  {author} {\bibfnamefont {G.}~\bibnamefont {Rosenman}},\ }\href {\doibase
  http://dx.doi.org/10.1063/1.123408} {\bibfield  {journal} {\bibinfo
  {journal} {Appl.\ Phys.\ Lett.}\ }\textbf {\bibinfo {volume} {74}},\ \bibinfo
  {pages} {914} (\bibinfo {year} {1999})}\BibitemShut {NoStop}%
\bibitem [{\citenamefont {K{\"o}nig}\ and\ \citenamefont
  {Wong}(2004)}]{Konig_Extended_2004}%
  \BibitemOpen
  \bibfield  {author} {\bibinfo {author} {\bibfnamefont {F.}~\bibnamefont
  {K{\"o}nig}}\ and\ \bibinfo {author} {\bibfnamefont {F.~N.~C.}\ \bibnamefont
  {Wong}},\ }\href {\doibase http://dx.doi.org/10.1063/1.1668320} {\bibfield
  {journal} {\bibinfo  {journal} {Appl.\ Phys.\ Lett.}\ }\textbf {\bibinfo
  {volume} {84}},\ \bibinfo {pages} {1644} (\bibinfo {year}
  {2004})}\BibitemShut {NoStop}%
\bibitem [{\citenamefont {Dorenbos}\ \emph {et~al.}(2008)\citenamefont
  {Dorenbos}, \citenamefont {Reiger}, \citenamefont {Akopian}, \citenamefont
  {Perinetti}, \citenamefont {Zwiller}, \citenamefont {Zijlstra},\ and\
  \citenamefont {Klapwijk}}]{dorenbos_superconducting_2008}%
  \BibitemOpen
  \bibfield  {author} {\bibinfo {author} {\bibfnamefont {S.~N.}\ \bibnamefont
  {Dorenbos}}, \bibinfo {author} {\bibfnamefont {E.~M.}\ \bibnamefont
  {Reiger}}, \bibinfo {author} {\bibfnamefont {N.}~\bibnamefont {Akopian}},
  \bibinfo {author} {\bibfnamefont {U.}~\bibnamefont {Perinetti}}, \bibinfo
  {author} {\bibfnamefont {V.}~\bibnamefont {Zwiller}}, \bibinfo {author}
  {\bibfnamefont {T.}~\bibnamefont {Zijlstra}}, \ and\ \bibinfo {author}
  {\bibfnamefont {T.~M.}\ \bibnamefont {Klapwijk}},\ }\href {\doibase
  http://dx.doi.org/10.1063/1.3003579} {\bibfield  {journal} {\bibinfo
  {journal} {Appl.\ Phys.\ Lett.}\ }\textbf {\bibinfo {volume} {93}},\ \bibinfo
  {eid} {161102} (\bibinfo {year} {2008})}\BibitemShut {NoStop}%
\bibitem [{\citenamefont {Boulanger}\ \emph {et~al.}(1998)\citenamefont
  {Boulanger}, \citenamefont {F\`{e}ve}, \citenamefont {Marnier},\ and\
  \citenamefont {M\'{e}naert}}]{Boulanger_Methodology_1998}%
  \BibitemOpen
  \bibfield  {author} {\bibinfo {author} {\bibfnamefont {B.}~\bibnamefont
  {Boulanger}}, \bibinfo {author} {\bibfnamefont {J.~P.}\ \bibnamefont
  {F\`{e}ve}}, \bibinfo {author} {\bibfnamefont {G.}~\bibnamefont {Marnier}}, \
  and\ \bibinfo {author} {\bibfnamefont {B.}~\bibnamefont {M\'{e}naert}},\
  }\href {http://stacks.iop.org/0963-9659/7/i=2/a=014} {\bibfield  {journal}
  {\bibinfo  {journal} {Pure Appl.\ Opt.}\ }\textbf {\bibinfo {volume} {7}},\
  \bibinfo {pages} {239} (\bibinfo {year} {1998})}\BibitemShut {NoStop}%
\bibitem [{\citenamefont {Pack}\ \emph {et~al.}(2004)\citenamefont {Pack},
  \citenamefont {Armstrong},\ and\ \citenamefont
  {Smith}}]{Pack_Nonlinear_2004}%
  \BibitemOpen
  \bibfield  {author} {\bibinfo {author} {\bibfnamefont {M.~V.}\ \bibnamefont
  {Pack}}, \bibinfo {author} {\bibfnamefont {D.~J.}\ \bibnamefont {Armstrong}},
  \ and\ \bibinfo {author} {\bibfnamefont {A.~V.}\ \bibnamefont {Smith}},\
  }\href {\doibase 10.1364/AO.43.003319} {\bibfield  {journal} {\bibinfo
  {journal} {Appl. Opt.}\ }\textbf {\bibinfo {volume} {43}},\ \bibinfo {pages}
  {3319} (\bibinfo {year} {2004})}\BibitemShut {NoStop}%
\bibitem [{\citenamefont {Saleh}\ and\ \citenamefont
  {Teich}(2013)}]{SalehTeich_Photonics_2013}%
  \BibitemOpen
  \bibfield  {author} {\bibinfo {author} {\bibfnamefont {B.}~\bibnamefont
  {Saleh}}\ and\ \bibinfo {author} {\bibfnamefont {M.}~\bibnamefont {Teich}},\
  }\href {http://books.google.com/books?id=Qfeosgu08u8C} {\emph {\bibinfo
  {title} {Fundamentals of Photonics}}},\ Wiley Series in Pure and Applied
  Optics\ (\bibinfo  {publisher} {Wiley},\ \bibinfo {year} {2013})\BibitemShut
  {NoStop}%
\bibitem [{\citenamefont {Zhong}\ \emph {et~al.}(2009)\citenamefont {Zhong},
  \citenamefont {Wong}, \citenamefont {Roberts},\ and\ \citenamefont
  {Battle}}]{Zhong_HighPerformance_2009}%
  \BibitemOpen
  \bibfield  {author} {\bibinfo {author} {\bibfnamefont {T.}~\bibnamefont
  {Zhong}}, \bibinfo {author} {\bibfnamefont {F.~N.~C.}\ \bibnamefont {Wong}},
  \bibinfo {author} {\bibfnamefont {T.~D.}\ \bibnamefont {Roberts}}, \ and\
  \bibinfo {author} {\bibfnamefont {P.}~\bibnamefont {Battle}},\ }\href
  {\doibase 10.1364/OE.17.012019} {\bibfield  {journal} {\bibinfo  {journal}
  {Opt.\ Express}\ }\textbf {\bibinfo {volume} {17}},\ \bibinfo {pages} {12019}
  (\bibinfo {year} {2009})}\BibitemShut {NoStop}%
\bibitem [{dix()}]{dixon_max-ent_2014}%
  \BibitemOpen
  \href@noop {} {}\bibinfo {note} {This is true for maximally entangled
  bipartite qubit states. The use of different states such as non-maximally
  entangled ones, can reduce this threshold.}\BibitemShut {Stop}%
\bibitem [{\citenamefont {Guha}\ and\ \citenamefont
  {Erkmen}(2009)}]{Guha_Gaussian_2009}%
  \BibitemOpen
  \bibfield  {author} {\bibinfo {author} {\bibfnamefont {S.}~\bibnamefont
  {Guha}}\ and\ \bibinfo {author} {\bibfnamefont {B.~I.}\ \bibnamefont
  {Erkmen}},\ }\href {\doibase 10.1103/PhysRevA.80.052310} {\bibfield
  {journal} {\bibinfo  {journal} {Phys. Rev. A}\ }\textbf {\bibinfo {volume}
  {80}},\ \bibinfo {pages} {052310} (\bibinfo {year} {2009})}\BibitemShut
  {NoStop}%
\end{thebibliography}

%

\end{document}